\documentclass{egpubl}
\usepackage{eg2017}

\STAR                   
\electronicVersion 

\ifpdf \usepackage[pdftex]{graphicx} \pdfcompresslevel=9
\else \usepackage[dvips]{graphicx} \fi

\PrintedOrElectronic

\usepackage{t1enc,dfadobe}

\usepackage{egweblnk}
\usepackage{cite}
\usepackage{hyperref}
 \usepackage{array,multirow,graphicx}
\usepackage[usenames,dvipsnames]{xcolor}
\usepackage{caption}
\usepackage{wasysym}
\usepackage{gensymb}
\usepackage{rotating}

\newcommand{\sylvain}  [1]{}
\newcommand{\cino}		[1]{}
\newcommand{\michi}    [1]{}
\newcommand{\jaiko}     [1]{}
\newcommand{\stefano} [1]{}
\newcommand{\jonas} [1]{}

\newcommand{\com}     [1]{{}}

\newcommand{\yes}     	{$\CIRCLE$}
\newcommand{\no}			{$\Circle$}

\title[From 3D Models to 3D Prints : an Overview of the Processing Pipeline]%
      {From 3D Models to 3D Prints:\\an Overview of the Processing Pipeline\\(preprint)}

\author[M. Livesu, S. Ellero, J. Mart\'{i}nez, S. Lefebvre \& M. Attene]
		{M. Livesu$^{1}$, S. Ellero$^{2}$, J. Mart\'{i}nez$^{3}$, S. Lefebvre$^{3}$ and M. Attene$^{1}$
        \\
         $^1$IMATI-CNR, Genova, Italy\\
         $^2$STAM SRL, Genova, Italy\\
         $^3$Inria, France\\
       }

\begin{document}


\maketitle

\begin{abstract}
Due to the wide diffusion of 3D printing technologies, geometric algorithms for Additive Manufacturing are being invented at an impressive speed. Each single step, in particular along the Process Planning pipeline, can now count on dozens of methods that prepare the 3D model for fabrication, while analysing and optimizing geometry and machine instructions for various objectives. This report provides a classification of this huge state of the art, and elicits the relation between each single algorithm and a list of desirable objectives during Process Planning. The objectives themselves are listed and discussed, along with possible needs for tradeoffs. Additive Manufacturing technologies are broadly categorized to explicitly relate classes of devices and supported features. Finally, this report offers an analysis of the state of the art while discussing open and challenging problems from both an academic and an industrial perspective.


\end{abstract}


\section{Introduction}
\label{sec::intro}
Digital 3D models have a central role in modern product development, and are becoming more and more important as Additive Manufacturing (AM) technologies are taking up in industrial practice. An industrial product has a typical lifecycle constituted of the four successive phases of \emph{conception}, \emph{design}, \emph{realization}, and \emph{service}: once a product \emph{concept} is designed by a CAD expert, the resulting 3D model must be realized through proper fabrication tools, and then put into service for actual use. Moving from design to realization amounts to defining a proper manufacturing method, that is, selecting  parts of the product (e.g. screws, bolts, bearings, ...) to be bought from external producers that offer a catalogue of standardized components, choosing and tuning fabrication tools for all the other parts, defining the machining instructions for such tools. All these operations are phases of the so-called \emph{Process Planning} (PP).
Thanks to their flexibility, the use of additive manufacturing technologies make the process planning a significantly easier task when compared with traditional methods (e.g. \textit{subtractive} manufacturing, moulding).

\subsection{Survey contents and objectives}
The objective of this report is to give the reader a comprehensive overview of the Process Planning pipeline in the context of Additive Manufacturing. 
We provide a deep picture of the algorithms, data-structures and shape representations that are involved in turning an input virtual model into a set of instructions that the AM machine can understand. We cover key publications from the infancy of AM to the most recent advances. We categorize algorithms based on the PP problem they solve, and discuss their impact on various desirable properties of the final parts.

\paragraph*{Process Planning only.} 
Recent surveys cover AM in a much broader sense, from a user's perspective. In particular, the reader can refer to the excellent survey by Gao and colleagues that proposes an overview of the status and challenges of AM in engineering~\cite{gao2015status}. It covers fundamental principles and available technologies, design and modelling for AM, and its impact on the industry and society at large.
The survey by Medeiros and colleagues~\cite{Medeiros:2016:SAF} provides a clear picture of recent contributions in \textit{functional} fabrication that is, parts that embed a functionality (e.g. articulated or deformable).
We do not cover these topics, or only mention them from the perspective of process planning.
%
We also do not cover topics regarding the design of the part. The reader can refer to \cite{hallgren2016re,thompson2016design,salonitis2015redesign} and references therein for recent discussions on the new trends in shape design for AM. Furthermore, we do not provide in depth technical reviews of the current 3D printing technologies. In Section~\ref{sec:technologies} we briefly go through the most important AM paradigms available. Our goal, however, is not to provide a comprehensive list, but rather to discuss the main properties -- and limitations -- that impact the process planning pipeline. 

\paragraph*{Motivation and timeliness.} 
Researchers operating in the area of process planning have been quite active in recent years, producing a tremendous amount of publications. 
The last (and only) survey focusing on the process planning for AM was published sixteen years ago \cite{kulkarni2000review} and, although the process planning pipeline has not changed much from an ideal point of view, many advances in algorithms have been proposed since. The same goes for surveys that focus on a more specific subject, such as \cite{dolenc1994slicing} and \cite{mohan2003slicing}, that review slicing procedures for AM and were published in 1994 and 2003, respectively. One exception is the recently released survey on the role of build orientation in layered manufacturing \cite{taufik2013role} which, however, reviews contributions coming from the mechanical engineering community, and does not cover recent literature coming from other fields (e.g. computer graphics).

\paragraph*{Target reader.} Indeed, additive manufacturing poses problems that cross several fields. Among classical players, such as designers, mechanical engineers and material scientists, new communities have become interested in it. For instance researchers in computer graphics, with their expertise at creating and manipulating digital 3D objects; but also applied mathematicians, with their expertise in fields such as topology optimization which is becoming increasingly important in the generation of support structures and infills. This report aims to be an entry door for those who want to contribute to AM, avoiding to re-invent the wheel, and promoting more interdisciplinary approaches. Both industry people and academics are targeted. 
From an industry perspective, this survey wishes to be a support to answer questions such as: I have a design, which  algorithms are suitable to process it? What geometric processes would I need to implement? 
From an academic perspective, scholars can use this report to validate the originality of their own ideas in the area, to look for existing solutions to similar problems, and to know which open challenges deserve their attention.


\subsection{General introduction to process planning}
Process Planning (PP) is the set of operations performed after designing a part or an assembly and before the actual manufacture of the components. PP techniques have been mostly developed in traditional subtractive manufacturing (SM) such as machining, and it is worth having a clear picture of these methods before entering the realm of modern Additive Manufacturing. Indeed, if in AM most of the steps can be performed algorithmically, in SM the process can be so complex that the experience of a skilled manufacturer is unavoidable.
Typically, such an expert must first analyse the part in order to identify the most suitable machining operation, namely: turning, milling, broaching, drilling/boring, etc. The selection of the most appropriate method mainly depends on the geometry of the part: cylindrical components require turning at least for the production of the overall body, components with flat or irregular surfaces must be milled, features such as slots and holes (standard, threaded, tapered, etc.) require specific operations such as broaching, drilling, tapping, boring, etc., grinding is needed for finished surfaces. Every machining operation has several configurations that may force the manufacturer to choose specific machines. For instance, milling machines can have 3, 4 or 5 axes, depending on the degrees of freedom of the part with respect to the tool or vice versa. Therefore, depending on the complexity of the part and/or of its features (undercuts, sloped surfaces, freeform surfaces, etc.), an appropriate machine is selected; a single part is often machined in several steps, occurring in different machines (e.g. turning, then milling, drilling and finally grinding).

The next step of the process is the definition of the stock, that is the piece of raw material to be machined. Obviously, the stock is selecting basing on the material of the part, on its overall dimensions and on the characteristics of the machine (e.g. only cylindrical parts can be loaded on lathes). For CNC machines, the defined stock, together with the characteristics of the machine (e.g. cutting speed, forwarding speed, maximum tool displacements, etc.) are input to a CAM software tool. It allows the simulation of each step of the machining process, together with the metrics of the process. The most important output of the process is the generation of the instructions for the CNC machine. The instructions, collected in a G-code file or similar format, are fed to the machine and include all the information about the toolpath, cutting parameters, the motion of the tool and of the part, lubrication and refrigeration of the part, etc.

Finally, the stock is placed in the machine and the zero position is set with reference to it; the G-code instructions are loaded and the actual manufacture of the part can start.

Innovative additive manufacturing processes have revolutionised the whole pipeline of manufacturing; process planning is the step which has benefited the most from this innovation. In fact, except for part finishing, the whole manufacture occurs in one machine. The user does not have to define any stock and also the toolpath computation is considerably simpler.

The shape of the part is not fabricated as a whole: it is manufactured as a union of layers (or slices), and layers are built one by one, one on top of the other, to form the final geometry. This involves a reduction of the PP problem dimension, because 2D toolpaths are generated (within each slice) instead of complex 3D paths. On the other hand, the result is now orientation dependent: gravity plays an important role in the manufacture of each layer. Moreover, its influence must be kept into account in the production of overhangs (which may require external supports), as stresses might be induced to the layer, potentially bringing to the failure of part production. 

\subsection{3D printing technologies}
\label{sec:technologies}

In this Section we briefly describe the main additive manufacturing technologies. Our goal is not to provide an extensive list, but rather to discuss the main properties -- and limitations -- that impact the process planning pipeline.

All the technologies we consider build an object layer after layer. They mainly differ, however, by whether they actually \textit{locally deposit} material or whether they \textit{solidify material} within an otherwise non-solid substance.
This property has important implications on the processing pipeline and this distinction therefore is the basis of our two main categories, \textit{material deposition} and \textit{layer solidification}.

A second fundamental distinction between these technologies is whether they deposit/solidify material along continuous paths (vector) or whether they rely on a discrete device (raster). This directly drives whether the output of the processing pipeline is a set of continuous paths (vector) or a set of images (raster).

\subsubsection{Material deposition}

Material deposition refers to methods that create the next layer by \textit{locally} depositing material on a previously printed layer. This encompasses techniques such as material extrusion (e.g. fused filament deposition~\cite{stratasys}), material jetting (e.g. UV sensitive resin droplets \cite{polyjet,multifab}), directed energy deposition (e.g. laser cladding~\cite{cladding}).

\paragraph{Vector or raster} Filament fused deposition is a vector approach as it deposits continuously along paths. The motion of the extruder is achieved through either a three axis gantry, or a delta robot configuration. Processes relying on resin droplets are usually discrete, similarly to inkjet printers. The print head is often attached to a two axis gantry, with the build plate moving up or down along the layering direction.

\paragraph{Properties} The key advantages of material deposition are the ability to combine multiple materials, a printing time that mostly depends on the part volume, and the ability to fully enclose voids. A major inconvenient however is the strong requirement for support structures, since material can only be deposited on top of an already existing layer. The process planning therefore has to automatically generate disposable support structures, which we discuss in Section~\ref{sec:sup:external}.

Finally, some of these technologies are able to print \textit{out-of-plane}, for instance generating a continuous spiralling path from bottom to top (e.g. \textit{spiralize} feature of the \textit{Cura} open source slicer) or even wire-frame structures \cite{Mueller:2014:WDP}. We discuss this in more details in Section~\ref{sec:slicing:alternatives}

\subsubsection{Layer solidification}

Layer solidification refers to all the processes that build the object by solidifying the top (or bottom) surface of a non-solid material (powder, liquid), typically within a tank. Layer fabrication starts by lowering the tank, adding a full layer of non-solid material, and then using a process that solidifies the material in specific places. This encompasses technologies such as vat photo-polymerization (e.g. stereolithography or SLA), powder bed fusion (e.g. selective laser sintering or SLS), binder jetting (e.g. plaster powder binding \cite{zcorp}), and sheet lamination (e.g. paper layering--cutting \cite{mcor}).

\paragraph{Vector or raster} SLA processes have both variants, relying either on a laser beam (vector) or on a projected image using a DLP projector (raster).
SLS processes are typically driving a laser beam through continuous motions, following contour paths. 

Both for SLS and SLA, beam motions are obtained using mirrors and galvanometer mechanisms -- forming a so-called laser scanner -- providing fast and precise movements.

\paragraph{Properties} A major advantage of layer solidification is the reduced need for support structures on complex geometries, enabling a much wider range of parts to print without any support. Note that supports may still be necessary to stabilize the part (see Section~\ref{sec:sup:external}) as it may be able to move within the non-solidified material (in particular with liquid resins, but also in powder depending on part weight). Another need for support arise from heat dissipation issues, in particular with metal powder melting.

A major drawback of within-layer solidification is the inability to mix different materials. This is  mitigated on some technologies by locally depositing additives, such as pigmented inks on powder binding 3D printers \cite{zcorp}. Another drawback is the necessity for non-solidified materials to exit cavities: this prevents the formation of fully closed empty voids within the part.

Finally, these approaches present an interesting tradeoff regarding printing times. Each layer starts by a full-surface layer filling which always takes the same, constant time. This implies that printing time is much more impacted by the height (number of layers) of the object, than by the solidified volume. As a consequence, printing a single small object is generally time consuming, while printing objects in batches can lead to significantly reduced print time per-parts, as the constant per-layer time is amortized. We discuss printing in batches in Section~\ref{sec::batch}.



\begin{table*}
\centering
\begin{tabular}{|l|c|c|c|c|c|}
\cline{2-6}
\multicolumn{1}{l|}{}&&&&&\\
\multicolumn{1}{l|}{}& \textbf{Vector/} & \textbf{Multiple}   & \textbf{Support} &  & \textbf{Build} \\
\multicolumn{1}{l|}{}& \textbf{Raster}   & \textbf{Materials} & 	\textbf{Structures} & \textbf{Cavities} & \textbf{Time} \\
\multicolumn{1}{l|}{}&&&&& \\
\hline
									& 				& 							& 							&							& \\
									& 				& 							& 							&							& Depends on\\
Material deposition		& Both 		& Supported			& Overhangs		& Supported$^*$			& part volume \\
									&  			& 							& 							& 							& \\
									& 				& 							& 							&							& \\
									&            & 		Not				& Stability/ 			 	& Not 					& Depends on \\
Material solidification	& Both		&  Supported		& 	Heat dissipation 	& supported			& part height	\\
									&            & 							& 							  	& 							& (number of layers)\\
\hline
\end{tabular}\\
\vspace{0.2cm}
$^*$: assuming the cavity can be printed without support structures.
\caption{Relation between broad AM technological solutions and supported features.}
\end{table*}

\subsection{Pipeline}
\subsubsection{Overview}
The traditional process planning pipeline starts from design specifications: these define not only the geometry of the part to be printed, but also additional requirements such as dimensional tolerances, need for surface finishing, materials to be used, and many other characteristics of the part.
Industrial practice in this phase is still dominated by 2D technical drawings, even if software tools (e.g. modellers from Autodesk and Dassault) and standard format specifications (e.g. STEP ISO 10303) exist to produce and represent accurate designs directly in 3D: in this latter case the so-produced models are known as CAD (Computer-Aided Design) models. Either 2D technical drawings or 3D CAD models must normally be cast to a corresponding CAM (Computer-Aided Manufacturing) representation to undergo the fabrication process, though some integrated CAD/CAM systems are emerging to make this transition as transparent as possible.

\begin{figure*}[tbh]
\centering
 \includegraphics[width=\linewidth]{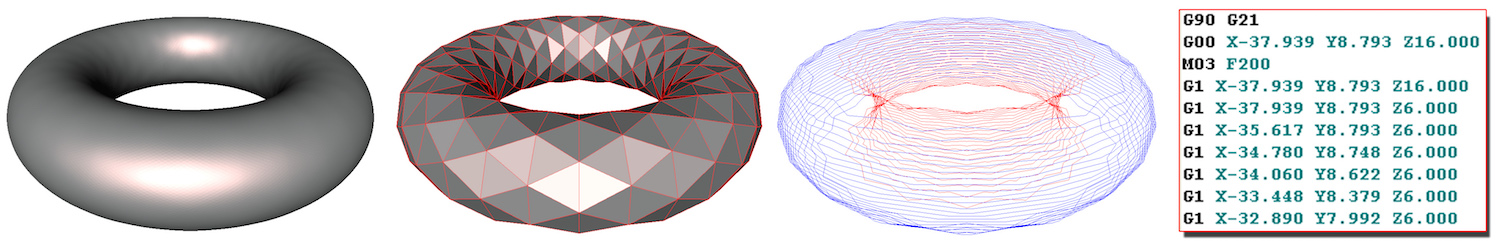}
\caption{Typical process planning phases in additive manufacturing: a design model (left) is tessellated to enter the Process Planning phase (centre-left). Such a tessellation is sliced (centre-right), and each slice is converted to a sequence of machine instructions (right). Other operations are typically required depending on the specific technology at hand.}
\label{fig:PP-overview}
\end{figure*}

Thus, in a standard product development pipeline, moving from the CAD to the CAM world represents the switch from the design to the process planning phase. Based on the CAM model, the final goal of Process Planning is to determine the machining instructions that the fabrication tool must execute to build the part.
Due to the layer-by-layer nature of Additive Manufacturing, it is important to select an appropriate building direction and to slice the model accordingly.
Each of the slices must then be converted to a proper \emph{toolpath}, that is, to a sequence of movements that the building tool must follow to fabricate the slice: these movements track the outer boundary of the slice, but also its inner parts and possible support structures. Figure \ref{fig:PP-overview} summarizes these main steps.

It is worth mentioning that process planning is typically an iterative procedure, and it might happen that the design specifications are not compatible with the fabrication technology (see Section~\ref{sec::fabreqs}): in this case the control must go back to the design phase for the necessary updates.
Fortunately, Additive Manufacturing technologies pose much fewer constraints on the design geometry (wrt to, e.g., CNC milling with limited degrees of freedom), but still there are cases where the aforementioned re-design cannot be avoided.

\subsubsection{Current practice}
Modern 3D printing companies and services accept designs in different representations, each leading to a different complexity for the conversion to an effective CAM model. In mechanical engineering, the vast majority of design models come as a collection of NURBS patches with possible trimming curves: such a \emph{nominal geometry} is normally tessellated to form a CAM model and start the process planning phase. When the shape is simple enough, constructive solid geometry (CSG) tools are normally preferred because they guarantee that the resulting model is solid. In this latter case, most CAD softwares provide a tessellation module, though a new trend of tools is emerging to completely avoid the tessellation and perform the whole process planning on the native representation, including the slicing phase.
In other areas (e.g. cultural heritage), the input model is usually produced by a 3D digitization campaign which often leads to a triangular mesh: such a representation is already tessellated, though many operations might be necessary to make it actually enclose a printable solid (see Section~\ref{sec::repairing}).

Whatever the format of the design geometry, the dominating format for the tessellated models used in CAM is the STL. STL represents an unstructured collection of triangles but is mostly used to represent structured meshes: thus, the coordinates of any vertex are encoded once for each of the triangles incident at that vertex, which leads to a highly redundant representation. Furthermore, STL files describe only the surface geometry without any representation of colour, texture or other common CAD model attributes.
For these reasons, modern standardization efforts (e.g. the AMF format within ISO/ASTM 52915:2013) use indexed representations to avoid redundancy and allow encoding many useful attribute information such as colours, materials and textures.


Nowadays, however, a typical AM process planning pipeline includes the following steps:
\begin{itemize}
\item \emph{Check and possible adaptation of the input geometry to fabrication requirements}. The (possibly tessellated) geometry must enclose a solid that the target printing technology can actually fabricate (Section~\ref{sec::pipeline});
\item \emph{Building direction}. The model must be correctly oriented to fit the printing chamber, minimize surface roughness and printing time, reduce the need for support structures, ... (Section~\ref{sec::orient});
\item \emph{Creation of support structures}. Depending on both the fabrication tool and the shape, additional geometry may be necessary to support overhanging parts and to keep the part from moving during printing (Section~\ref{sec::supports});
\item \emph{Slicing}. The model must be converted to a set of planar slices whose distance might be either constant or adaptive (Section~\ref{sec::slicing});
\item \emph{Machine instructions}. Each slice must be converted to either a sequence of movements of the fabrication tool (vector-based) or a grid of pixels that define the solid part of the slice (raster-based) (Section~\ref{sec:mach_instr}).
\end{itemize}





\section{Metrics / Desiderata}
\label{sec::desiderata}

Setting up the process planning to fabricate an object is a matter of finding a good tradeoff between 
different objectives, often depending on the applicative scenario. The PP pipeline can be tuned to strive to optimize 
for one, or a combination of them. Different criteria have been proposed in the literature. We recap here the most 
relevant and widely used.

\subsection{Cost}
\label{sec:desiderata:cost}
In industrial environments it is quite important to keep the production cost as low as possible. The Generic Cost Model \cite{alexander1998part} puts together all the variables that control the production cost for a single object, and is designed to be general enough to embrace any layered manufacturing process. Similar cost models have been proposed in \cite{byun2006determination,Byun2006bis,thrimurthulu2004optimum,pham1999part,xu1999considerations}. The typical cost model is defined as the sum of three major components: pre-build, build, and post-processing.

\subsubsection{Pre-build cost} 
Measures the cost necessary to turn a design into a set of machine instructions to send to the printer. It also accounts for the labour cost (e.g. load the powder into the machine, control or supervise the process planning software) and the time necessary to setup the printer (e.g., cleaning, testing, warming). Methods that aim to minimize the pre-build cost strive for the efficiency of the process, which can be achieved either by using computationally more efficient algorithms or by reducing the user interaction, favouring automatic methods.

\subsubsection{Build cost}
Comprises the material cost (for both the part and the supports) and the cost of using the machine for the time necessary to complete the job. The build cost can be reduced in two ways: acting on the printing time, or acting on the material waste. Printing time can be reduced orienting the shape so as to minimize its height (Section~\ref{sec::orient}),
reducing the number of slices (Section~\ref{sec::slicing}), or using efficient machine toolpaths (Section~\ref{sec:mach_instr}). Material can be reduced  by minimizing the support structures' volume, either with a proper choice of the build direction (Section~\ref{sec::orient}) or by inserting cavities in the interior of the shape \cite{Song-2016-CofiFab,Lu:2014:BTL,Wang:2013:CEP}. Recent works aim to reduce material waste and achieve a better surface finish by splitting the part into components that can be printed without supporting structures at all (Section~\ref{sec:multi:qual}). Notice that these methods often trade minimal build cost for the structural strength of the part and, therefore, are not always suitable for industrial production processes.

\subsubsection{Post-processing Cost}
\label{sec:post_processing_cost}
Measures the labour, material and time cost necessary to polish the part. This includes: detaching the support structures from the object at the end of the print (Section~\ref{sec:sup:external}), and applying some surface finish technique, either manually or through chemical and machine driven processes. These components heavily depend on the process plan. Particularly relevant is the choice of the build direction (Section~\ref{sec::orient}) which, in turn, determines the amount and positioning of support structures (Section~\ref{sec:sup:external}), and the extent to which the staircase effect introduced by the layered manufacturing process will affect the part quality.




\subsection{Fidelity}
\label{sec:desiderata:fidelity}
\begin{figure}[h]
\centering
 \includegraphics[width=\linewidth]{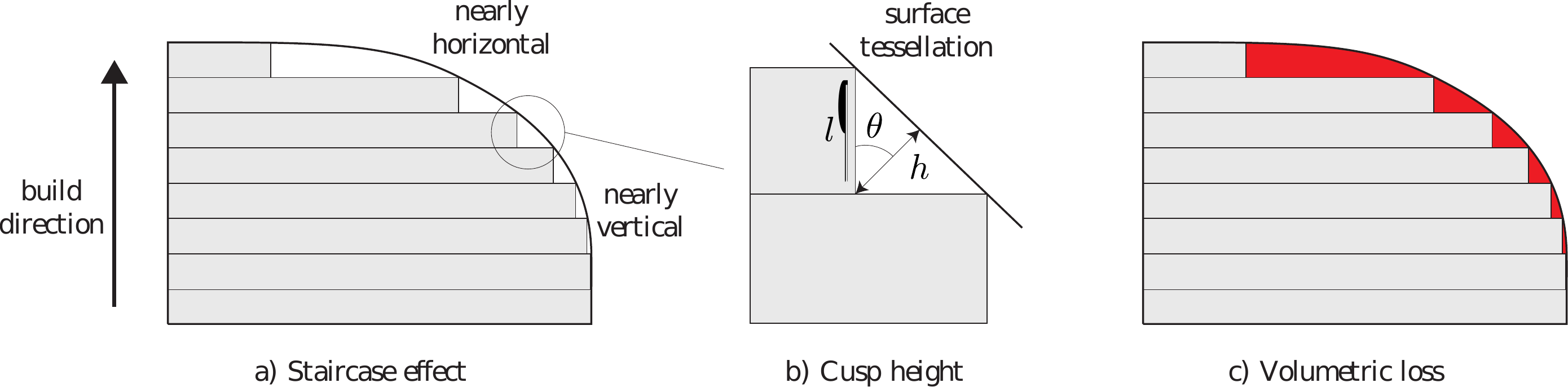} 
\caption{Approximating a curved surface with a stack of layers piled along the building direction introduces the typical staircase effect and reduces fidelity (a). As can be noticed nearly horizontal surfaces introduce more error than nearly vertical ones. Cusp height (b) and volumetric difference (c) are among the most widely used proxies to estimate fidelity.}
\label{fig:staircase_error}
\end{figure}

Is the degree of exactness with which the part has been reproduced starting from its design. Indeed, layered manufacturing is hardly capable of producing a perfect replica of a given design. Parts of the shape that do not align with the building direction expose a typical \emph{staircase} effect (Figure~\ref{fig:staircase_error}a). We distinguish between \emph{form} and \emph{texture}, where the former refers to the overall shape of the prototype, which, to a certain extent, can be quantitatively estimated before printing, and the latter to more local variations of the surface (or high frequencies), which can be approximately estimated only on the printed object since they depend on factors like the printer resolution and the material used. In Section~\ref{sec:quality_metrics} we discuss the most widely used metrics to evaluate the form approximation error introduced by the staircase effect. Being dependent only on process plan parameters, such as the layer thickness and the shape orientation, these metrics are at the core of the algorithms that strive to optimize the PP pipeline (see e.g. Section~\ref{sec::orient}). In Section~\ref{sec:surface_finish} we introduce metrics to evaluate surface texture, and also discuss some of the practices used in literature to alleviate surface artifacts. Finally, in Section~\ref{sec:design_compliancy} we briefly discuss design compliancy, a criterion that is of fundamental importance in industrial applications, where the printed object is asked to stay within the tolerances set by the designer.

\subsubsection{Form}
\label{sec:quality_metrics}
In additive manufacturing objects are created by approximating a given design with a set of layers stacked one on top of the other along the build direction. The goal of form fidelity is to detect the difference in shape between these two entities. We present here the two most widely adopted metrics in literature to evaluate the error introduced by layered manufacturing: the cusp height error and the volumetric difference.

\begin{itemize}
\item The \emph{cusp-height error} concept was introduced in the context of adaptive slicing for layered manufacturing \cite{dolenc1994slicing}. It is defined as the maximum distance between the manufactured part's surface and the design surface \cite{alexander1998part}. It depends on the layer thickness $l$ and the angle $\theta$ between the local surface orientation and the build direction (Figure~\ref{fig:staircase_error}b), namely
\begin{equation}
h = 
\left\lbrace 
\begin{array}{cc}
l \vert \cos \theta \vert	  & \quad\textbf{for }\quad  \vert \cos \theta \vert \neq 1\\
\\
0  & \quad\textbf{for }\quad  \vert \cos \theta \vert = 1
\end{array}
\right.
\end{equation}
Note that $\vert \cos \theta \vert $ is very small for nearly orthogonal angles and grows up to $1$ when $\theta$ is close to $0$ degrees. This well encodes the big approximation error difference between nearly vertical and nearly horizontal surfaces, as shown in Figure~\ref{fig:staircase_error}a. The integral of the cusp height on the whole surface is a good estimator of the fidelity of the printed model. To this end, for triangulated surfaces the cusp height is computed separately on each facet as the dot product between the build direction $b$ and the triangle normal $n$, $\vert \cos \theta \vert = \vert b\cdot n \vert $. It is then scaled by the triangle area and normalized by the total mesh area so as to accommodate uneven tessellations and make it scale independent \cite{Wang:cgf:improved};\\

\item The \emph{volumetric difference} is the difference between the volume enclosed by the design and the volume of the printed object (red area in Figure~\ref{fig:staircase_error}c). In \shortcite{masood2000part} Masood and colleagues explain how to evaluate the volumetric error for simple geometries such as cylinders, cubes, pyramids and spheres. Such work was then extended to complex geometries in \cite{masood2003generic}. 
\end{itemize} 

Notice that, although they are both used as proxies to minimize the staircase effect, volumetric difference and cusp height are not equivalent. Taufik and colleagues \shortcite{taufik2014volumetric} express a preference for volumetric error. They observe that, if the shape to be printed contains steep slopes, to a little variation in cusp height may correspond a large variation in volumetric difference, thus making the latter more accurate than the former.

\subsubsection{Texture}
\label{sec:surface_finish}
Also known as \emph{surface roughness}, or \emph{surface finish}, texture aims at measuring tiny local variations on the surface, which determine visual (e.g. the way the object reflects the light), haptic (e.g. the porosity of the surface) and mechanical (e.g. friction) properties of the shape. Unlike form, texture cannot be estimated prior printing, as it mostly depends on parameters like the printer resolution and the printing material. Usually it is measured directly on the printed object, by using sampling techniques \cite{townsend2016surface}.

In industrial environments the design specifies both the desired surface roughness and the metric that should be used to estimate it. The arithmetic mean surface roughness ($Ra$) is by far the most widely adopted metric, as stated in \cite{townsend2016surface}. Delfs and colleagues \shortcite{Delfs2016} observe that the surface roughness depth ($Rz$) is a better proxy to measure surface finish, as it is well representative of how a human eye assesses surface quality. Both $Ra$ and $Rz$ are roughness metrics defined in the ISO 4287 standard \cite{standard4287}. They are computed on profile curves obtained by cutting the surface with an orthogonal plane. We point the reader to \cite{roughness_buletin} for a nice explanation of how these metrics are defined and can be estimated in practice; much other information can be found online. 

Different strategies have been proposed in the literature to achieve the best surface quality possible. In \shortcite{reeves1997reducing} Reeves and Cobb evaluate the \emph{meniscus smoothing} to alleviate the impact of the staircase effect in stereolithography and produce smoother surfaces. It consists of an edited build cycle in which each layer, after solidification, is lift above the upper surface of the resin tank to stretch a meniscus of liquid between each polymerized layer. The resin meniscus is then solidified by using scan data from the previous layer, producing a smoother transition between adjacent layers. Other authors have recently observed that additional artifacts that affect surface finish may be introduced while detaching support structures from the object. In fact, tiny features may be too weak and break during this process, leaving residual support material attached to the surface. Zhang and colleagues \shortcite{zhang:sa:2015} propose a perceptual method that optimizes for the location of the touching points between surface and supports. Their system tries to \emph{hide} support removal artifacts by placing them at the least salient parts of the shape, as far as possible from its perceptually relevant features. In general, the extent to which support removal may affect surface quality depends on the printer and the material used. In metal printing supports removal is extremely challenging due to the properties of the material involved. Some recent methods to alleviate this problem are discussed in Section~\ref{sec:sup:external}.

\subsubsection{Design compliancy}
\label{sec:design_compliancy}
In industrial AM environments objects are requested to be compliant with the original design, meaning that both the form and the texture must stay within precise error bounds set by the designer. To this end, a number of form and orientation tolerances are often used in industrial design. Typical form requirements regard the straightness, planarity, circularity or cylindricity of the components. Regarding orientation, typical requirements are parallelism, orthogonality or angularity. Notice that, similarly to roughness, these tolerances can only be estimated after printing. Metrology is a vast field. We do not discuss here details regarding how this quantities can be estimated. We point the reader to \cite{townsend2016surface,huang2015quality,jin2015out,huang2014predictive} for further details on this topic.

\subsection{Functionality}
When the object to be printed is the result of a shape optimization process, it is typically asked to meet some prescribed functionality requirements. We consider three broad categories: requirements on the robustness of the shape, such as resiliency with respect to previously known or unknown external forces; requirements on the mass distribution, for example to achieve static or dynamic equilibrium; and requirements on thermal and mechanical properties, for example regarding the heat dissipation or the stiffness of industrial components.

\subsubsection{Structural soundness / Robustness}
\begin{figure}[t]
\centering
\includegraphics[height=2.8cm]{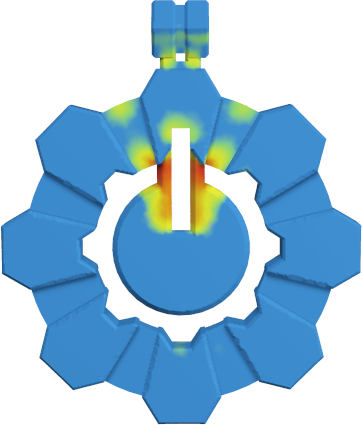}
\includegraphics[height=2.8cm]{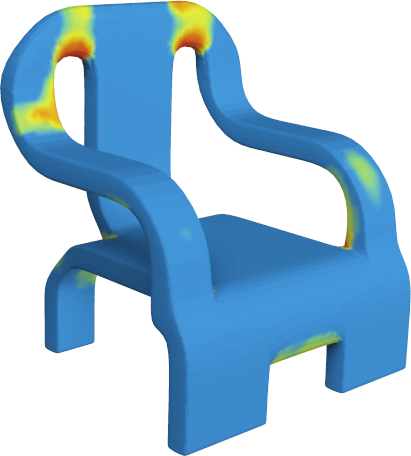}
\includegraphics[height=2.8cm]{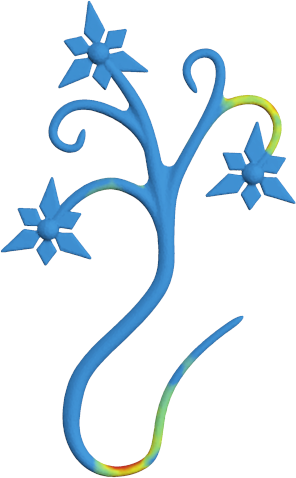}
\includegraphics[height=2.8cm]{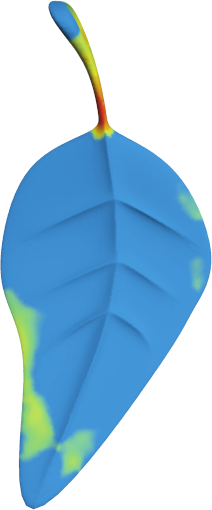}
\caption{Structural analysis to detect weak object parts. Image courtesy of \shortcite{zpz_struct_ays_13}.}\label{fig:worst_case_analysis}
\end{figure}

With 3D printing consumers can directly produce their own objects, but not always the digital shapes they started with were meant to be fabricated, and can therefore reveal to be excessively fragile and easily break under cleaning, transportation or handling. Zhou and colleagues \shortcite{zpz_struct_ays_13} introduced the worst case structural analysis to detect the most fragile parts of an object (see Figure~\ref{fig:worst_case_analysis}). 
In \cite{langlois:2016:SSA} a stochastic finite element method to compute failure probabilities is presented.
A number of methods  \cite{Xu2016,li2015interior,stava.12.sig} combine a lightweight structural analysis with automatic systems that enforce weak features through operations such as hollowing, thickening, strut insertion and inner structures.

\subsubsection{Mass distribution}
\label{sec:design_compliancy:mass}
The distribution of material and cavities inside a 3D printed model has been the subject of recent research. In \cite{Prevost:MIS:2013} a method to optimize the balance of a shape to make it stand in a given pose is proposed. The work was further extended in \cite{SpinIt:Baecher:2014}, where a novel optimization of the mass distribution to make an abject spinnable around a given axis was proposed. The optimization of the buoyant equilibrium is the subject of \cite{wang:eg:2016}, where a method to create floating objects in a prescribed pose is presented. In \cite{Prevost:BMM:2016} the authors consider the standing, suspension, and immersion balancing problems for 3D printed objects containing embedded movable masses. In \cite{wu2016shape,musialski2015reduced} unified frameworks for the optimal interior design and mass distribution of 3D printed objects that enable the optimization of static, rotational and buoyant equilibrium are presented. Recent research has pushed this concept even further.  In \cite{musialski_2016_sosp} the shape design is posed as non-linear problem that aims to optimize the natural frequencies of a shape, for example to make it sound in a controlled manner, as a musical instrument.

\subsubsection{Thermal / Mechanical properties}
Additive manufacturing enables the fabrication of shapes that would be impossible to produce with classical subtractive techniques. Shapes that in the past were interesting only from a pure theoretical standpoint can now be printed and their functionality exploited. To this end, additive manufacturing has fostered a lot of research in fields like topology optimization \cite{dede2015topology,brackett2011topology}, where the goal is to generate shapes which optimize performances in terms of some physical requirement (e.g. weight, heat dissipation, stiffness). This is important in fields like aerospace industry, where the size and weight of components is a crucial factor, or for the generation of injection moulds, where optimized cooling systems may increase both the productivity and the overall quality of industrial products.
Alongside, researchers have been investigating the correlation between the structure of a shape and its physical properties.
Recent works aim to optimize the local structure of a printed object to produce controlled deformations or to meet precise rigidity requirements. We review this body of literature in Section~\ref{sec:sup:internal:microstruct}.



\section{Process Planning steps}
\label{sec::pipeline}

\subsection{Meeting fabrication requirements}
\label{sec::fabreqs}
Herewith we distinguish between requirements of the shape and requirements of its representation. Shape requirements are printer-specific, and define rules for the compatibility of the geometry with the printing hardware. Representation requirements guarantee that the (tessellated) geometry to be printed actually encloses a solid without ambiguity.

\subsubsection{Shape requirements}

\paragraph{Checks} When the input model comes in raster form (e.g. a voxelization), \cite{telea11_ismm} provides the means to analyze the shape and identify problematic regions whose size drops below the printing resolution (e.g. thin walls or other tiny features). In a slightly more general setting, \cite{xavier2013thickness} describes an approach to estimate the thickness of triangulated models.
\paragraph{Automatically fixing the input}
 While the previous works are meant to \emph{detect} possible incompatibilities, in \cite{wang2013thickening} an algorithm is proposed to actually thicken sheet-like structures so as to make them printable. In \cite{stava.12.sig} a similar approach is presented with a focus on the structural characteristics of the printed prototype: note that this method is not meant to make the model printable, but it shares several aspects with the previous one. If the model cannot fit into the printing chamber due to its size, \cite{luo2012chopper} proposes an approach to split the model into parts that can be printed separately and reassembled after printing. Apparently no publication deals with the automatic placement of drainage channels for models with internal cavities to be printed with powder bed technologies.

\subsubsection{Input representation requirements}
\label{sec::repairing}
Geometry repairing has received increased attention in recent years, not only for 3D printing, but in general for all the scenarios where a "well-behaving" mesh is required (e.g. Finite Element Analysis, advanced shape editing, quad-based remeshing, ...). Some repairing methods transform the input into an intermediate volumetric representation and construct a new mesh out of it \cite{taoju04} \cite{kobbelttog} \cite{Chen:2013:RCG}. In a new trend of methods specifically tailored for 3D printing, a 3D mesh is converted into an implicit representation, and all the subsequent operations (including the slicing) are performed on this representation \cite{Huang:2013:IFA} \cite{icesl}. These methods are very robust but necessarily introduce a distortion.
Robustness and precision are indeed major issues in this area, in particular when self-intersections must be removed \cite{attene2014}. In this case some approaches rely on exact arithmetics \cite{nefpolyhedron}, while some others can losslessly convert the input into a finite precision plane-based representation, and then reconstruct a provably good fixed mesh out of it \cite{campen10} \cite{wang13}.
When used for 3D printing applications, however, the aforementioned exact approaches are useful only if the input actually encloses a solid, while they are not really suitable to fix meshes with visible open boundaries \cite{attene10}.
For a more comprehensive overview of mesh repairing methods, we point the reader to \cite{attene2013} and \cite{taoju09}.

\begin{figure}[htb]
	\centering
	\includegraphics[width=.99\linewidth]{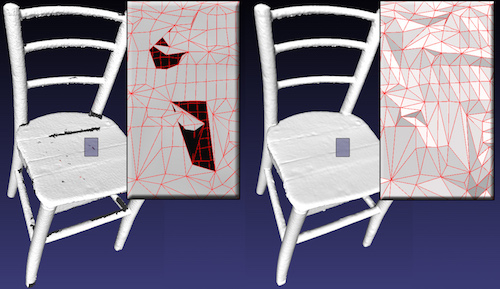}
	\caption{A raw digitized mesh may not enclose a solid due to various defects (left). Mesh repairing algorithms perform little modifications that turn such a raw model to an actual polyhedron that bounds a solid without ambiguity (right). Image courtesy of \cite{attene10}.}
	\label{fig:repaired_chair}
\end{figure}

Since 3D printing can only produce solid objects, a repairing algorithm must ensure that the resulting mesh actually encloses a solid \ref{fig:repaired_chair}. If the input mesh has open boundaries, a typical solution is to fill the holes in advance and then rely on some of the previously mentioned repairing methods. This approach is employed by one of the most popular web-based mesh fixing services \cite{netfabb}, but it makes sense only if the boundaries are actually delimiting surface holes. Unfortunately, in some cases the designer uses zero-thickness surfaces to represent sheet-like features (e.g. a flag), and for models of this kind a hole-filling approach would produce rather coarse results. Another widely used software that performs mesh repairing for 3D printing is Autodesk's Meshmixer \cite{meshmixer}, where the input STL can be successfully fixed even if it has open boundaries but at the cost of an overall approximation due to the global remeshing approach employed.

\subsection{Orientation}
\label{sec::orient}

The choice of the building direction is crucial in layered manufacturing as it directly influences the time necessary to 
print the object, the amount of support structures necessary to sustain the part during the print and the surface quality \cite{taufik2013role,pandey2007part,alexander1998part}. The first algorithms to select a proper part orientation date back to the mid 1990's \cite{allen1994computation,richard1995optimizing,frank1995expert,cheng1995multi,lan1997determining,hur1998development}. Several other methods have been proposed ever since, each one striving to optimize either for one, or a combination of the criteria discussed in Section~\ref{sec::desiderata}.

Directly optimizing for the build direction in the space of all possible orientations is often too complex due to the non smooth nature of the metrics involved \cite{ezair2015orientation}. Former approaches used to consider a small number of candidate orientations, either predefined or computed on a shape proxy (e.g. the convex hull). Many recent methods (e.g. \cite{Wang:cgf:improved,morgan2016part,ezair2015orientation}) share a similar heuristic:
they start with a regular sampling of the possible orientations; shortlist the orientations that perform best according to some quality metric, and, starting from each of them, navigate the space of solutions looking for the closest local minimum. The building orientation is eventually defined as the one corresponding to the lowest of the local minima explored by such heuristic.

Table~\ref{table:orient} summarizes the properties of the orientation optimization techniques we discuss next.

\subsubsection{Optimize for Cost}
\label{sec:orient_cost}
Recent works attempt to optimize for the orientation of the part using the volume of the support structures as only metric. Ezair and colleagues \shortcite{ezair2015orientation} showed that the resulting function is continuous but non-smooth with respect to the orientation angles. In \cite{ezair2015orientation,khardekar2006fast} two GPU-based volume estimation of supports structures are proposed. Morgan and colleagues \shortcite{morgan2016part} define the support volume as the sum of the volumes of the prisms generated by extruding the down-facing triangles up to the building plate.

\subsubsection{Optimize for Fidelity}
\label{sec:orient:fidelity}
Delfs and colleagues \shortcite{Delfs2016} propose an orientation system that optimizes for surface roughness, using the mean roughness depth as a proxy to optimize for surface finish. One of the key features of the proposed approach is the ability to prescribe local accuracy requirements, attaching a target surface finish to each triangle in the tessellation. Masood and colleagues \shortcite{masood2003generic,masood2000part} proposed two systems that aim to find the building orientation that minimizes the volumetric error (Section~\ref{sec:desiderata:fidelity}). 

In \cite{hildebrand2013orthogonal} the authors address the problem of finding the proper build orientation for objects to be printed at low resolution via laser cut (cardboard or plywood). Their contribution is the definition of an optimal orthonormal frame that is suited for the decomposition into smaller parts, each of which to be sliced along one of the three directions with small volume loss error.

Other works focus on the artifacts that may be generated when detaching support structures. In \cite{ahn2007fabrication} the authors investigate how to orient the shape so as to minimize post processing (i.e. supports removal and surface finish). In \shortcite{zhang:sa:2015} Zhang and colleagues introduced a perceptual model to find preferable building directions in 3D printing, so as to place support structures in the least salient parts of the object.

\subsubsection{Optimize for Functionality}
\label{sec:orient:functionality}
Due to the layered nature of the process, the build orientation can significantly affect the performance of the resulting objects, introducing structural anisotropy. Ulu and colleagues \shortcite{ulu2015enhancing} propose a FEM-based building orientation optimization that maximizes the minimum factor of safety (FS) under prescribed loading and boundary conditions. Umetani and colleagues \shortcite{umetani2013cross} propose a cross-sectional structural analysis based on bending momentum equilibrium. This is used in particular for orientation optimization. Their method avoids computationally expensive FEM simulations, and can be plugged into interactive modeling tools to allow users consider structural robustness during incremental trial-and-error design.

\subsubsection{Optimize for mixed factors}
In \cite{phatak2012optimum} and \cite{thrimurthulu2004optimum} two genetic algorithms to find the optimal part orientation are presented. They both formulate the subject of their optimization as a weighted sum of multiple quality criteria, regarding both cost and fidelity. The weights can be finely tuned by the user to set tune the importance of each criterion. Similar approaches have been presented in \cite{Byun2006bis,byun2006determination}, however these methods do not scale well with complex mechanical shapes (and non-mechanical shapes) as they consider a very restricted set of candidate orientations computed on the convex hull of the part.

Decision support systems to aid RP users chose the best building direction according to their needs are presented in \cite{pham1999part} and \cite{hur1998development}. Both systems consider multiple criteria that can be prioritized according to the user needs, such as: overhang area, supports volume, build time and cost. On the negative side, these systems are specialized for CAD shapes and do not scale well on free-form shapes.

%


\begin{table*}[ht]
\centering
\begin{footnotesize}
\begin{tabular}{|l|cc|ccc|c|}
\hline
\multicolumn{1}{|c|}{\textbf{}} & \multicolumn{6}{c|}{\textbf{}}\\
\multicolumn{1}{|c|}{\textbf{METHOD}} & \multicolumn{6}{c|}{\textbf{OPTIMIZES FOR}}\\
\multicolumn{1}{|c|}{\textbf{}} & \multicolumn{6}{c|}{\textbf{}}\\
 & \multicolumn{2}{c|}{\textbf{Cost}} & \multicolumn{3}{c|}{\textbf{Fidelity}} & \multicolumn{1}{c|}{\textbf{Functionality}}\\
& Slice number  & Supports volume 	& Cusp height    &Volume loss 	& Surface finish 	& Stress resiliency\\
\hline
\cite{Delfs2016}								&	\no	&	\no	&	\no	&	\no	&	\yes	&	\no\\ \hline
\cite{morgan2016part}					&	\no	&	\yes	&	\no	&	\no	&	\no	&	\no	\\ \hline
\cite{Wang:cgf:improved}				&	\no	&	\yes	&	\yes	&	\no	&	\no	&	\no	\\ \hline
\cite{ezair2015orientation}				&	\no	&	\yes	&	\no	&	\no	&	\no	&	\no	\\ \hline
\cite{ulu2015enhancing}					&	\no	&	\no	&	\no	&	\no	&	\no	&	\yes	\\ \hline
\cite{zhang:sa:2015}						&	\no	&	\no	&	\no	&	\no	& \yes	&	\no	\\ \hline
\cite{umetani2013cross}				&	\no	&	\no	&	\no	&	\no	&	\no	&	\yes	\\ \hline
\cite{hildebrand2013orthogonal}	&	\no	&	\no	&	\no	&	\yes	& \no	&	\no	\\ \hline
\cite{phatak2012optimum}				& \yes	& \yes 	& \yes 	& \no 	& \yes 	& \no 	\\ \hline
\cite{ahn2007fabrication}				&	\no	&	\no	&	\yes	& \no	&	\yes	&	\no	\\ \hline
\cite{Byun2006bis}							&	\yes	&	\yes	&	\yes	&	\no	&	\no	&	\no\\ \hline
\cite{byun2006determination}		&	\yes	&	\yes	&	\yes	&	\no	&	\no	&	\no	\\ \hline
\cite{thrimurthulu2004optimum}	&	\yes	&	\yes	&	\yes	&	\no	&	\no	&	\no	\\ \hline
\cite{masood2003generic}				&	\no	&	\no	&	\no	& \yes	&	\no	&	\no	\\ \hline
\cite{masood2000part}					&	\no	&	\no	&	\no	& \yes	&	\no	&	\no	\\ \hline
\cite{pham1999part}						&	\yes	&	\yes	&	\yes	&	\no	&	\no	&	\no	\\ \hline
\cite{hur1998development}			&	\yes	&	\yes	&	\yes	&	\no	&	\no	&	\no	\\
\hline
\end{tabular}
\end{footnotesize}
\caption{Techniques for orientation optimization and their properties, see Section~\ref{sec::orient}.}
\label{table:orient}
\end{table*}



\subsection{Support Structures}
\label{sec::supports}

Support structures are a key component of process planning. They are used to compensate for some limitations of
the manufacturing processes, in particular maximum overhang angles beyond which deposited material falls, and the large increase in time due to printing inner volumes of a part.

In the following we categorize supports into two main categories: disposable \textit{external} supports that assist the fabrication process and are removed afterwards (Section~\ref{sec:sup:external}), and \textit{internal} supports that  modify the inside of the object to achieve a trade-off between material cost, print time and physical properties (Section~\ref{sec:sup:internal}).

\subsubsection{External}
\label{sec:sup:external}

\begin{figure}[t]
	\centering
	\includegraphics[width=\linewidth]{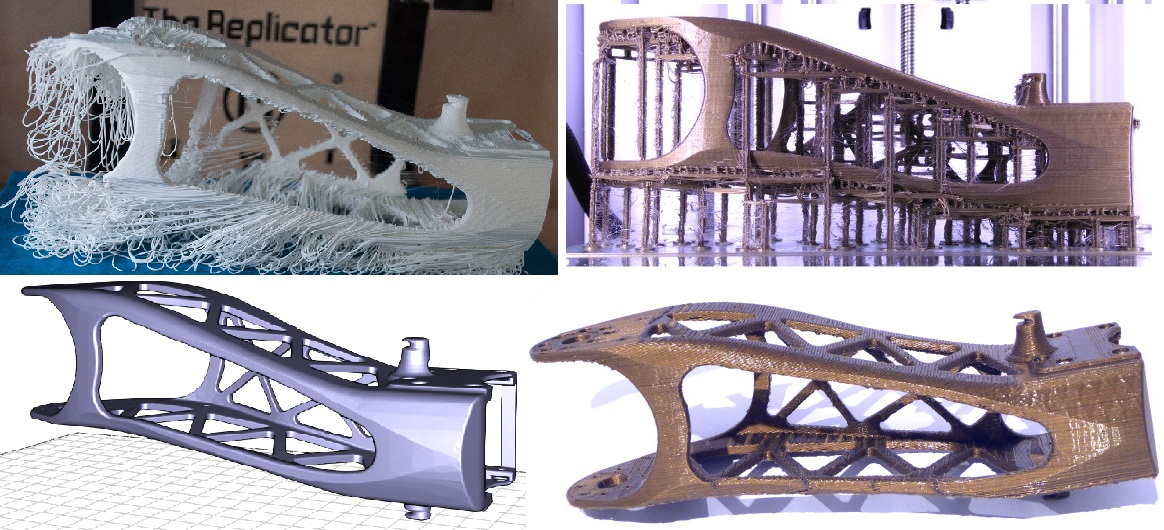}
	\caption{\textbf{Top left}: A robot upper leg (3D model below) printed without support. Filament falls due to excessive overhangs. \textbf{Top right}: A sacrificial external support structure, and the cleanup model (below). 
		{\small (3D model from the Poppy project \url{https://www.poppy-project.org/en/}, image from~\cite{dumas2014bridging})}.
	}
	\label{fig:extsupport}
\end{figure}

External support structures are sacrificial structures that are fabricated alongside the object. After fabrication completes they are chemically or mechanically removed. This usually involves human intervention and therefore is a time consuming, expensive step. An example of a complex part printed with and without support is shown in Figure~\ref{fig:extsupport}.

Part orientation is a major factor in external support requirements, and therefore is often optimized to reduce the need for supports (see Section~\ref{sec::orient}). Researchers have also proposed methods to slightly deform the design so has to reduce support~\cite{Hu20151} and even to design models that are guaranteed to print without any support~\cite{reiner.16.egshort}. However in most cases the process planner has to comply with the input model and some amount of supports remains necessary.

\noindent Different types of external supports serve different purposes:
\begin{itemize}
\item Local deposition technologies can only deposit material on existing surfaces below. Thus, surfaces appearing mid-air and surfaces at an excessive overhang require support just below them.
\item Shapes may move or deform during the fabrication process. This typically happens when fabricated objects are imbalanced and when the raw material (powder, resin) cannot sustain the weight of the print. Another source of distortion are stresses from thermal gradients. To reduce these issues supports acting as fixtures are necessary.
\item Some processes can generate a large quantity of heat, in particular metal printing. This excessive heat accumulation results in shape distortions and residual stresses. In such a case, additional supports may act as heat diffusers.
\end{itemize}

\noindent We next review approaches from the literature and discuss their use for each type of support. First, however, let us take a closer look
at when and why supports are required.

\paragraph*{Islands, overhangs, and self-supporting surfaces.}

\begin{figure}[t]
	\centering
	\includegraphics[width=\linewidth]{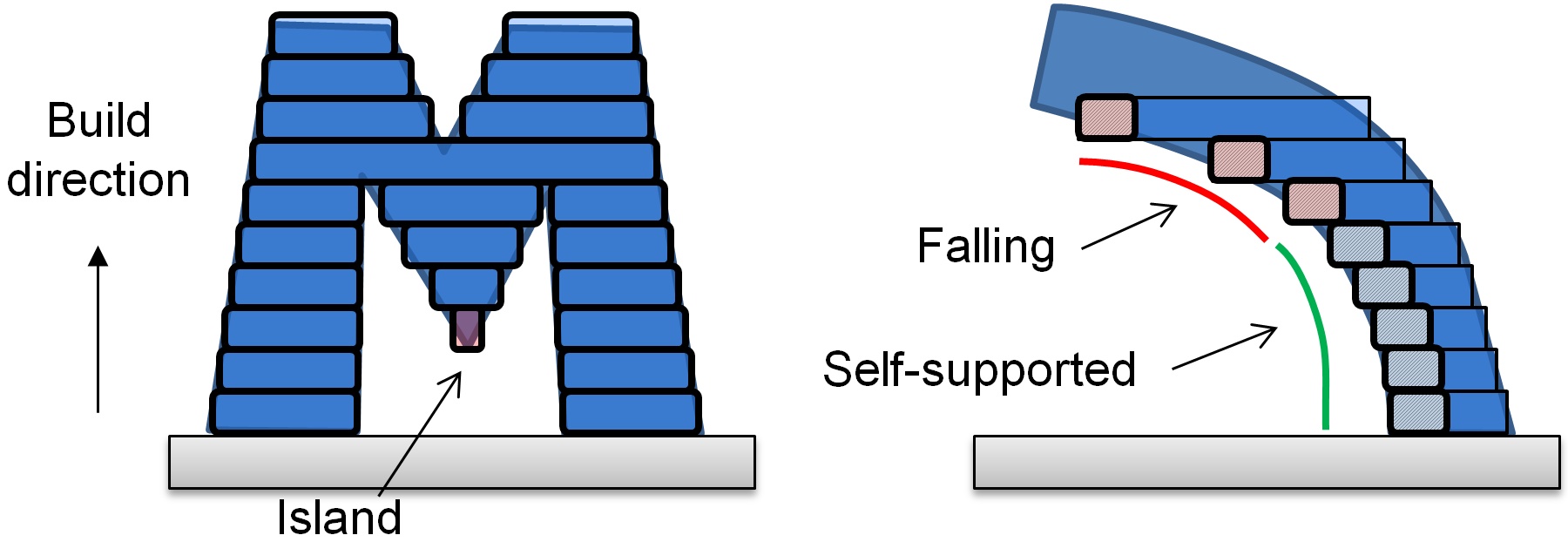}
	\caption{\textbf{Left}: The M letter is decomposed into layers. The downward facing tip becomes an island with respect to the build direction.
		     \textbf{Right}: Each layer is solidified from the outside towards the inside. The hatched rectangles are the cross sections of material deposition paths. After a critical angle, the deposited material no longer bonds to the layer below and falls.
	}
	\label{fig:supportcases}
\end{figure}

A first situation that requires support is illustrated in Figure~\ref{fig:supportcases}, left. After dividing the shape into layers, one of the slices
contains an island -- a solid region that appears while not being supported from below. During additive fabrication from bottom to top, the material forming
the island will not attach to already solidified material. For technologies using material deposition, the material forming the island will fall. As a consequence, material deposited onto the next layer above will also fall, and this will cascade into a catastrophic failure. For technologies using layer solidification, non solidified material below (e.g. powder) will usually be able to support the island -- however weight might accumulate over several layers and a heavy disconnected component may start to sink down. For technologies using resins (SLA), the island is problematic: it will typically end up floating in the viscous resin, cascading into a complete print failure. 

Overhangs also can produce problematic cases with material deposition. The material is typically added progressively along deposition paths. The right illustration of Figure~\ref{fig:supportcases} shows a cross section of the deposited paths for an overhang region. Due to the layering, at some excess overhangs the deposited material will simply fall (red rectangles in the Figure). It is interesting to note, however, that until some threshold angle the deposited material will have a sufficient bonding surface with the layer below. This self-supporting property stems from the bonding between successive layers and allows for overhangs to exist up to some maximum angle without requiring support.
Overhangs are less of an issue for technologies employing layer solidification -- even though excessive overhangs may distort due to change in material properties during solidification, or to auxiliary motion during preparation of the next layers. 

\paragraph*{Detecting surfaces requiring support.}
Generating supports for areas in overhang requires two main steps: the detection of the surfaces in need for support, and the generation of the support structure itself.
For detecting surfaces in overhang a first family of approaches consider the down-facing facets of the input mesh having an angle too steep to print correctly (e.g. ~\cite{kirschman.91.asme,allen1995determination}).
A second family of approaches consist in performing a boolean difference between two successive slices (e.g. ~\cite{patent.3dsystems.88,chalasani.95.support}). The width of the difference determines which regions are self-supporting~\cite{chalasani.95.support}. This can also be detected conveniently and efficiently with two-dimensional morphological operations in image space~\cite{huang.09.tst,Chen:2013:DGP,huang.chapter.14}. As discussed in~\cite{huang.chapter.14} care must be taken however with some specific configuration, where some protruding regions might be mis-classified as supported (see Figure 1.8 in this publication).
\cite{dumas2014bridging} performs the detection directly at the toolpath level, verifying whether each deposited segment is supported by at least half its width from below.

This first analysis generally leads to a compact set of points to be supported.
Several approaches then select a subset by down-sampling~\cite{patent.materialize.07,Chen:2013:DGP,dumas2014bridging,huang.chapter.14}.

\paragraph*{Generating a support structure.}
Once the surfaces requiring support are determined, the support geometry is computed. The main trade-off is between print time, material use, and reliability.

For instance, for filament printers the traditional approach consists in extruding the mesh facets requiring support downward, thus defining a large \emph{support volume}. The support volume is usually printed with a weak infill pattern (see also Section~\ref{sec:sup:internal:infills:sparse}). This still uses a significant amount of material and time, but it is very reliable: the support typically has a large area of contact with both the part and the print bed, ensuring the print stability in most cases.
In this context several approaches modify the support volume to reduce its size. Huang et al.~\shortcite{huang.09.ijamt} use sloped walls instead of straight walls for the sides, shrinking the support volumes in their middle sections. Heide \shortcite{patent.stratasys.10} reduce the support volume by decreasing its size and complexity as the distance below the supported model increases.

Other 3D printing technologies can print complex and thin structures more reliably. In the context of SLA, Eggers and Renap~\shortcite{patent.materialize.07} form a support structure by starting from a regular rhombus mesh filling the print bed. The 3D model is subtracted from the initial structure, removing intersected mesh edges. Points requiring support are attached to the mesh by downward angled beams.
Huang et al. \shortcite{huang.chapter.14} produce a support made of a sparse set of vertical pillars connected by angled beams for structural strength. The position of the pillars is optimized and their pairwise connections follows a minimal spanning tree (as seen from above) to keep the support structure small. 
\textit{MeshMixer}\cite{schmidt2014branching} builds a thin structure supporting the part in a sparse, limited number of points, generating a support structure resembling a tree. This approach has been used successfully on both FDM and SLA machines. Vanek et al. \shortcite{vanek2014clever} propose an algorithm to optimize for similar tree support structures.
Wang et al. \shortcite{Wang:2013:CEP} optimize truss structures for the primary purpose of strengthening 3D printed objects, and extend their approach for support generation. Support beams are added by tracing rays downwards.
Dumas et al. \shortcite{dumas2014bridging} generate bridge scaffoldings that rely only on vertical and horizontal bars, improving reliability and part stability while generating small support structures. Remarkably, the horizontal bars can be printed efficiently on FDM printers, but are also well suited for SLA. 
The static shape balance at all stages of the process is taken into account, enlarging the support structure whenever necessary.
Calignano et al.~\cite{calignano2014design} discuss the design of support structures for SLM (selective laser melting) where supports act both as fixtures and heat dissipater. 


\paragraph*{Support removal.}
External support structures are meant to be detached from the part after the print is completed. Depending on the materials involved (e.g. printing on metal) this operation can be extremely challenging. Furthermore, residual material can remain attached to the surface, badly affecting the surface finish. In polymer AM processes, to alleviate the supports detaching problem, thermoplastic materials that dissolve in alkaline baths are used \cite{priedeman2004soluble}. Hildreth et al. \shortcite{hildreth2016dissolvable} have recently shown that similar approaches are possible even for stainless steel printing, were the differences in the electrochemical stability between different metals can be exploited to dissolve carbon steel supports. Jhabvala et al. \cite{jhabvala2012innovative} exploited the pulsed laser radiation to print metal supports that are both faster to print and easier to remove; the system supports only SLM printers. A valid alternative consists in trying to orient the part in such a way that the supports necessary to sustain it during the print will stick only to the least salient portions of the shape, as proposed in \cite{zhang:sa:2015} (Section~\ref{sec:orient:fidelity}).


\subsubsection{Internal supports}
\label{sec:sup:internal}

The interior of an object is a key factor regarding the material use, print time and mechanical properties of the final result. 
The impact on material use and print time is easily explained by the fact that the inner volume grows to the cube of the scaling factor (i.e. doubling the size of an object multiplies its volume by eight). Therefore, most of the time and material is spent on the inside of the object. Carving the inside can lead to large savings. However, care must be taken to ensure that the object can still be fabricated (e.g. avoiding the introductions of overhangs or islands, not forming pockets) and that the end result will remain rigid enough. 
In addition, modifying the structure of the inside of an object gives the opportunity to change its global mechanical behaviour, for instance making it flexible or rigid in different places, or changing its balance.

We organize this section as follows. 
We first consider in Section~\ref{sec:sup:internal:hollow} techniques that focus on creating large empty pockets within objects. Next, we discuss in Section~\ref{sec:sup:internal:infills:dense} and in Section~\ref{sec:sup:internal:infills:sparse} approaches that focus on how to infill the object interior, from dense to sparse infill patterns. These techniques have an emphasis on material/time savings and exploit specificities of the processes. 
We focus in Section~\ref{sec:sup:internal:frames} on frame structures, which are typically beam or cellular structures optimized to create a strong structure within the object interior.
We discuss in Section~\ref{sec:sup:internal:microstruct} techniques that focus on changing the mechanical behaviour of the object by filling its inside with micro-structures. Since this is a very large topic, we keep the focus on techniques that tightly integrate with the process planning pipeline. 


\paragraph{Hollowing.}
\label{sec:sup:internal:hollow}

Most of the techniques we discuss here treat the general geometric problem of computing an inner cavity at a fixed distance from the object surface. The surface of the inner cavity is called the \emph{offset surface} of the model~\cite{farouki1985exact,rossignac1986offsetting}.
These approaches can be used to compute inner cavities that are either left empty or filled with the approaches described in Section~\ref{sec:sup:internal:infills:sparse}.

\begin{figure}[b]
\centering
\includegraphics[width=\linewidth]{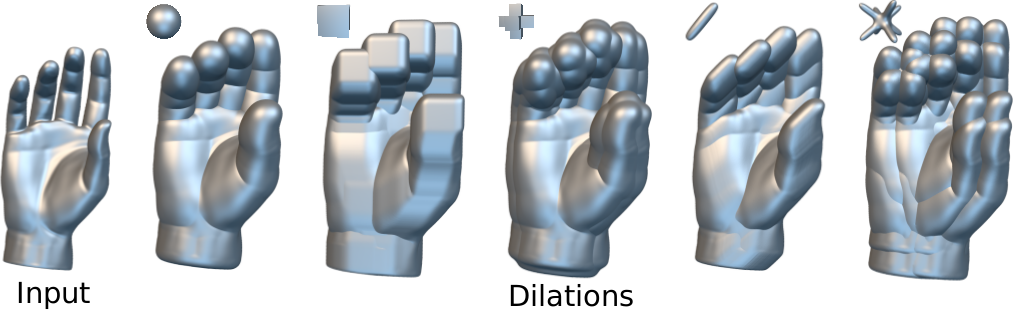}
\caption{Morphological dilation of an input point cloud, considering different structuring elements. Image courtesy of \shortcite{stephane2014point}.}\label{fig:point_morphology}
\end{figure}

Early approaches obtain a superset of geometric primitives of the offset surface that are trimmed and filtered to form the final offset boundary~\cite{forsyth1995shelling}. Qu and Stucker~\shortcite{qu2003surface} presented a vertex convolution method for STL files without explicit treatment of self-intersections. Campen and Kobbelt~\shortcite{campen2010polygonal} introduced an exact convolution approach.
%
Hollowing can also be performed by first computing the \textit{distance field} of the model, and extracting the offset surface from it.
Frisken et al.~\shortcite{frisken2000adaptively} presented and adaptively sampled distance field.
Varadhan and Manocha~\shortcite{varadhan2006accurate} approximate the offset surface with a distance field isosurface extraction, guaranteeing a Hausdorff distance bound on the approximation.
Pavi\'{c} and Kobbelt~\shortcite{pavic2008highresolution} traverse an octree distance field and split each cell which is potentially intersected by the offset surface.
Liu and Wang~\shortcite{liu2011fast} extract the isosurface of a narrow band distance field.

Another class of hollowing methods consider ray representations of solids such as the \emph{dexel structure}~\cite{Hook:1986:RTS} 
or the layered depth normal images (LDNI)~\cite{Chen:2008:LDN}.
For a single direction and a uniform grid of rays parallel to that direction, a ray-representation stores the
intervals of the rays lying inside the solid.
Hui~\shortcite{hui1994solid} computes the sweeping of a solid along a trajectory by considering the union of a finite set of ray representations of the solid.
Chiu and Tan~\shortcite{chiu1998using} computes the morphological erosion of each dexel, taking into account its neighbourhood dexels.
Hartquist et al.~\shortcite{hartquist1999computing} compute the union of spheres over the boundary of the input model.
Wang and Manocha~\shortcite{wang2013gpubased} place spheres on the samples of a LDNI structure, and compute their union efficiently on the GPU.
Chen and Wang~\shortcite{chen2011uniform} generate a convolution geometric primitives of the offset surface, constructs their LDNI, and filters the points of the superset that belong to the offset surface.
Mart\'{i}nez et al.\shortcite{martinez2015chained} consider the dilation of a dexel structure along different successive ray directions, and exploit the winding number of specially constructed meshes.
Other methods generate a voxelization of the offset surface. Li and McMains~\shortcite{li2014sweep} present a GPU approach to compute the Minkowski sum of a polyhedra by computing pairwise Minkowski sums, and obtain a voxelization of their union.
%
Hollowing can also be done by considering the offset of surface points.
Lien et al.~\shortcite{lien2008covering} computes the Minkowski sum between two surfaces sampled by points, and distinguish the interior and boundary points.
Calderon and Boubekeur~\shortcite{stephane2014point} introduced a set of morphological operations for point clouds (see Figure~\ref{fig:point_morphology}).

A few methods hollow the model during slicing, that is at slice level. 
McMains et al.~\shortcite{McMains:2000:LMO} considers regularized boolean operations of each slice contour, in order to approximate the offset surface.
In order to achieve uniform hollowing thickness, Park~\shortcite{park2005hollowing} considers the erosion of a circle swept over the slice contours. 

We focused here on hollowing at the process planning stage. As mentioned in Section~\ref{sec:design_compliancy:mass} hollowing can also be used
at design time to change the mass distribution of an object and optimize for various properties such as balance.



\paragraph{Dense infills.}
\label{sec:sup:internal:infills:dense}

Most technologies such as SLS, SLA, binder deposition and fused filament deposition support dense infilling. On raster devices, it suffice to produce an image of the filled layer contour. On vector devices, densely filling the object requires to follow a space filling curve when solidifying the material. The curve represents the path followed by the deposition device, while it deposits (or solidifies) a wide and thick track of material. The thickness matches the layer height, while the width depends on the technology (typical values are $20\mu m$ for focused lasers, $400\mu m$ for plastic extrusion). Thus, the spacing between two neighbouring paths has to match the deposition width. A smaller spacing produces excess deposition/solidification (\textit{overflow}) while a larger spacing leaves gaps (\textit{underflow}).
 
\begin{figure}[h]
	\centering
	\includegraphics[width=\linewidth]{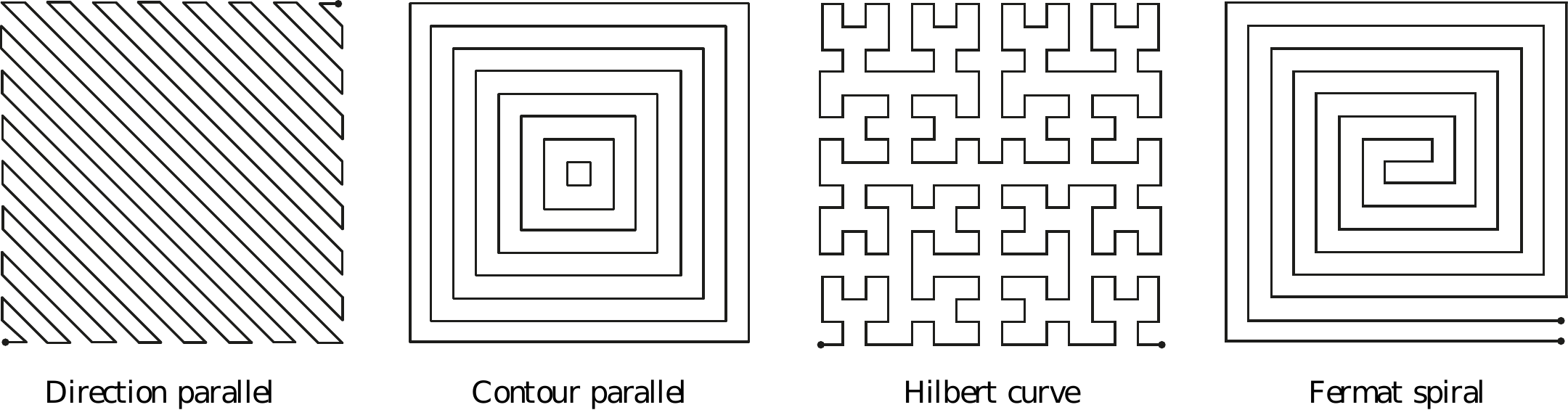} 
	\caption{Four different filling patterns used in additive manufacturing. From left to right: direction parallel, contour parallel, the Hilbert space filling curve and the recently proposed Fermat spiral.}
	\label{fig:toolpath_patterns}
\end{figure}

The shape of the space filling curve can have an impact on both the print time and the final object strength. Some possible patterns are illustrated in Figure~\ref{fig:toolpath_patterns}.

For most technologies -- including fused filament deposition -- a popular pattern is the so called \emph{direction parallel} (or \emph{zig-zag}) toolpath \cite{McMains:2000:LMO}, which consists in filling the slice area with a set of equally spaced segments parallel to one another and linked at one of their extremities. The spacing between the hatches is adjustable to control the final density, and the direction of the hatches changes every two slices to avoid strong mechanical biases. 
To gain speed or save material, the parameters used during infilling may be different, e.g. reducing binder flow, or changing the laser focus to solidify larger tracks.

{Contour parallel} infill \cite{yang2002equidistant,held1994pocket,kao1998optimal} is a common alternative to direction parallel infill. The contour parallel pattern consists in a set of concentric closed curves that emanate from the outer boundary of the slice and propagate inwards (see Figure~\ref{fig:toolpath_patterns}). These are in fact the offset contours from the layer boundary.
This is an appealing pattern since it closely follows the outer contour of the slices. Unfortunately it also tends to leave gaps within the slices. Indeed, the offset curves from the outer contour meet around the medial axis, and it is unlikely that the remaining space matches the deposition width. These gaps are difficult to handle. For this reason most slicers rely on a hybrid approach that combine direction parallel and contour parallel~\cite{McMains:2000:LMO}.
A few contour parallel curves are produced near the boundary of the slice, and the remaining inner polygon is filled the interior with a zig-zag pattern. The most critical part in the implementation of such hybrid approaches is the handling of the meeting point between the contour parallel and the direction parallel toolpaths, in which detaching, under and overfilling may occur~\cite{jin2013adaptive_other}. 

Infill patterns can lead to two issues: First, they may result in many stops and restarts of material deposition, which can lead to several problems (e.g. many small gaps, material feeder failure). Second, sharp turns may slow down the motion or introduce vibrations. 
For instance, direction parallel toolpaths can lead to many small segments if the slice contains thin walls orthogonal to the deposition direction -- inducing a large number of turns. 
The speed of different dense infill strategies have been compared, including variants that smooth out the sharp turns between hatches~\cite{kim2002machining,el2006toolpath}. To obtain a more continuous pattern, Ding et al.~\cite{ding2014tool} decompose the slice in different regions that use different directions of parallel infill. The patterns from one region to the next are smoothly connected. 
Zhao et al.~\cite{Zhao:2016:CFS} advocate for the use of spirals. A key advantage is to reduce the number of sharp turns thereby enabling faster motions, while achieving a fill pattern that resembles the contour parallel infill.

Other alternatives to these dominant patterns are Hilbert spiral \cite{griffiths1994toolpath}, Moore's spirals \cite{cox1994space}, Peano spirals \cite{makhanov2007advanced}. A common drawback for both spirals and contour parallel patterns is the lack of \emph{direction bias}, which prevents the generation of the cross weaved layouts used by direction parallel approaches. 



\paragraph{Sparse infills}
\label{sec:sup:internal:infills:sparse}

The cost of additive manufacturing being essentially driven by time and material use, sparse infilling is an important feature. The geometry of a sparse infill is often determined by the target process. In particular, powder and resin systems (e.g. SLS/SLA) cannot create closed voids: non solidified material is trapped and cannot exit, the sparsity would be lost. However, they can print geometries that are significantly more complex than, e.g. fused filament fabrication. This lead researchers to propose internal frame structures that we describe in Section~\ref{sec:sup:internal:frames}, as well as micro--structures described in Section~\ref{sec:sup:internal:microstruct}.

Fused filament fabrication (FFF) and similar technologies (e.g. contour crafting) require dedicated infill patterns, due to the strong overhang constraints. In addition, the thin slanted beams that are used with other processes are slow and less reliable to print. 
Most slicing software for FFF support sparse infills. These infills are usually 2D zig-zag hatching patterns that are vertically extruded~\cite{dinh.15.sigcourse}, again possibly changing orientation every two slices to avoid strong mechanical biases. 
\cite{Kumar:2009:FRT} explores hierarchical versions of space filling curves to grade the density of the infill pattern.

These patterns are efficient and simple to compute during slicing, and offer a good support for the roofs of the inner cavity. However, due to the vertical extrusion they are not mechanically strong if pressured on the sides.
To obtain stronger patterns, Steuben et al.~\shortcite{Steuben:2016:ISF} define the infill as the iso-contours of a scalar field, or the principal directions of a vector field within each slice. For instance, the infill paths can follow the principal directions of stress from a finite element simulation, resulting in stronger patterns for a given load scenario. 

\begin{figure}[t]
	\centering
	\includegraphics[width=.4\linewidth]{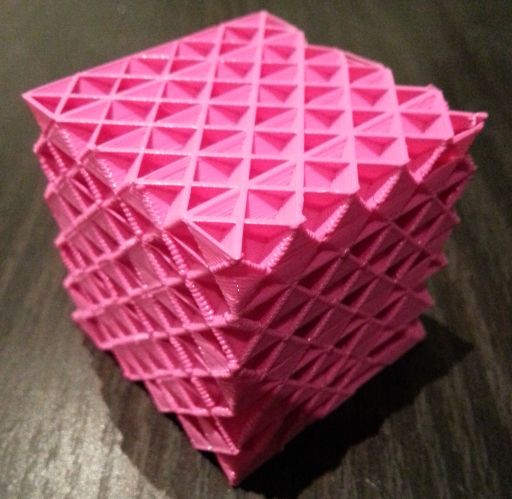} \hspace{5mm}
	\includegraphics[width=.371\linewidth]{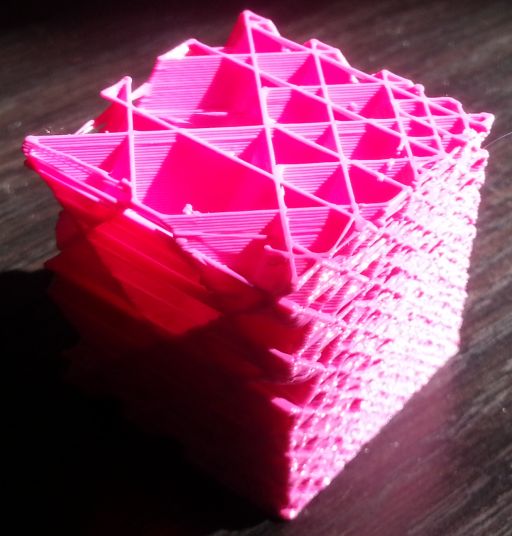} 
	\caption{Rhombic infill and its hierarchical version.}
	\label{fig:rhombic}
\end{figure}

There has also been several attempts to move beyond simple vertical extrusions. 
3D printing enthusiasts have experimented with interesting 3D infill patterns~\cite{nicoll.11.blog}. 
The software \textit{Slic3r} proposes an infill pattern that produces a 3D honeycomb pattern. 
This recently led to a new type of pattern called \textit{rhombic infill}. These are formed by the intersection of at least three sets of parallel planes in space. These infill have a number of interesting properties. First, they can be efficiently generated during the slicing process, and by carefully choosing the angles of the inner planes they can print as fast as vertical extrusions~\cite{lefebvre.15.blog}. Second, when printed with uniform density they are very strong thanks to the inner 3D cell structures. Third, they can be subdivided to locally increase the infill density, as shown in Figure~\ref{fig:rhombic}, right. Wu et al.~\shortcite{Wu:2016:SSR} propose a criteria based on the overall rigidity and balance of the object to perform this subdivision. Lee and Lee~\cite{Lee:2016:BBI} subdivide closer to the cavity roofs in order to create large empty cavities. They further reduces the size of the structure by removing  faces that are not required to support a structure above. 

In an attempt to make infill as sparse as possible, Hornus et al.~\shortcite{hornus.16.gradifab} propose a method that creates maximal inner carvings while ensuring that they remain fabricable with filament deposition (the cavities are self-supporting). This is achieved through morphological operations on the slices, growing a self-supporting inner cavity from the object tops.


\paragraph{Internal frame structures}
\label{sec:sup:internal:frames}
As mentioned earlier, frame structures are especially well suited for SLA/SLS technologies, even though they have also
been successfully demonstrated on fused filament fabrication.
Wang et al.~\shortcite{Wang:2013:CEP} propose to fill an object interior with a sparse truss structure. The structure is optimized to reduce the number of beams while preserving  rigidity.
Zhang et al.~\shortcite{Zhang:2015:MAT} similarly produce an inner structure made of beams but instead exploit the medial axis of the object, using it as a backbone structure. 
Medeiros et al.~\shortcite{Medeiros:2015:AV} generate an adaptive tessellation of the interior, and offset the edges of either the primal or the dual to produce an inner beam structure. Thanks to the adaptive tessellation, the structure is denser along the shape boundary than on the inside. 
Lu et al.~\shortcite{Lu:2014:BTL} optimize for a Voronoi diagram inside the print, which faces form an infill pattern. 

A difficulty in designing 3D infill patterns is that their complexity might lead to increased print time, for the same density (this is less true on systems such as SLA/DLP where the entire layer is exposed at once). Yaman et al.~\shortcite{Yaman:2016:SCI} consider how to print efficiently the faces of a Voronoi diagram, following an Euler cycle to solidify in sequence the segments forming the faces in each slice.


\paragraph{Microstructures}
\label{sec:sup:internal:microstruct}

Micro-structures are internal infill patterns that seek to change the macroscopic physical behavior of the final object. For instance, even when printing with a single material certain micro-structures modify the elastic behavior of the object, making it more or less flexible. It is often possible to grade and control the change across the final object.

The design of microstructures with tailored properties was introduced in the 1990's~\cite{Sigmund1995:S,cadman2013design}. A large range of techniques deal with optimizing functional microstructures, such as functionally graded materials for CAD applications~\cite{jackson1999modeling,kou2007heterogeneous,oxman2011variable} and porous scaffold design for bioengineering~\cite{hollister2005porous}, among many other applications. 
A complete review of the field falls out of the scope of this document. Instead, we focus here only on the techniques working in conjunction with the process planning. 
In particular, as the size of microstructures becomes smaller, approaches explicitly storing the microstructure geometry become computationally infeasible.

Chen~\shortcite{Chen:2007:DTM} defines micro-structures as periodic tiles that are then efficiently mapped into the volume, similarly to volume texture mapping. By deforming the mapping, the infill locally adapts to a density field. Pasko et al.~\shortcite{pasko2011procedural} considered procedural definitions of periodic microstructures. The parameters of the microstructures can vary spatially to produce graded materials~\cite{Fryazinov:2013:CAD}, for instance to reinforce an object following a cross-sectional stress analysis~\cite{li2015interior}.
OpenFAB~\cite{Vidimce:2013:OPP} provides a specialized language to describe procedural microstructures. The geometric details are efficiently evaluated at slicing time, streaming voxels to the printer.
The advantages of procedural representations for the process planning have been identified early by Park and colleagues in the context of multi-material fabrication~\cite{Park:2001:VMT} (see also Section~\ref{sec:multimat}).

\begin{figure}[t]
\centering
\includegraphics[width=0.3\linewidth]{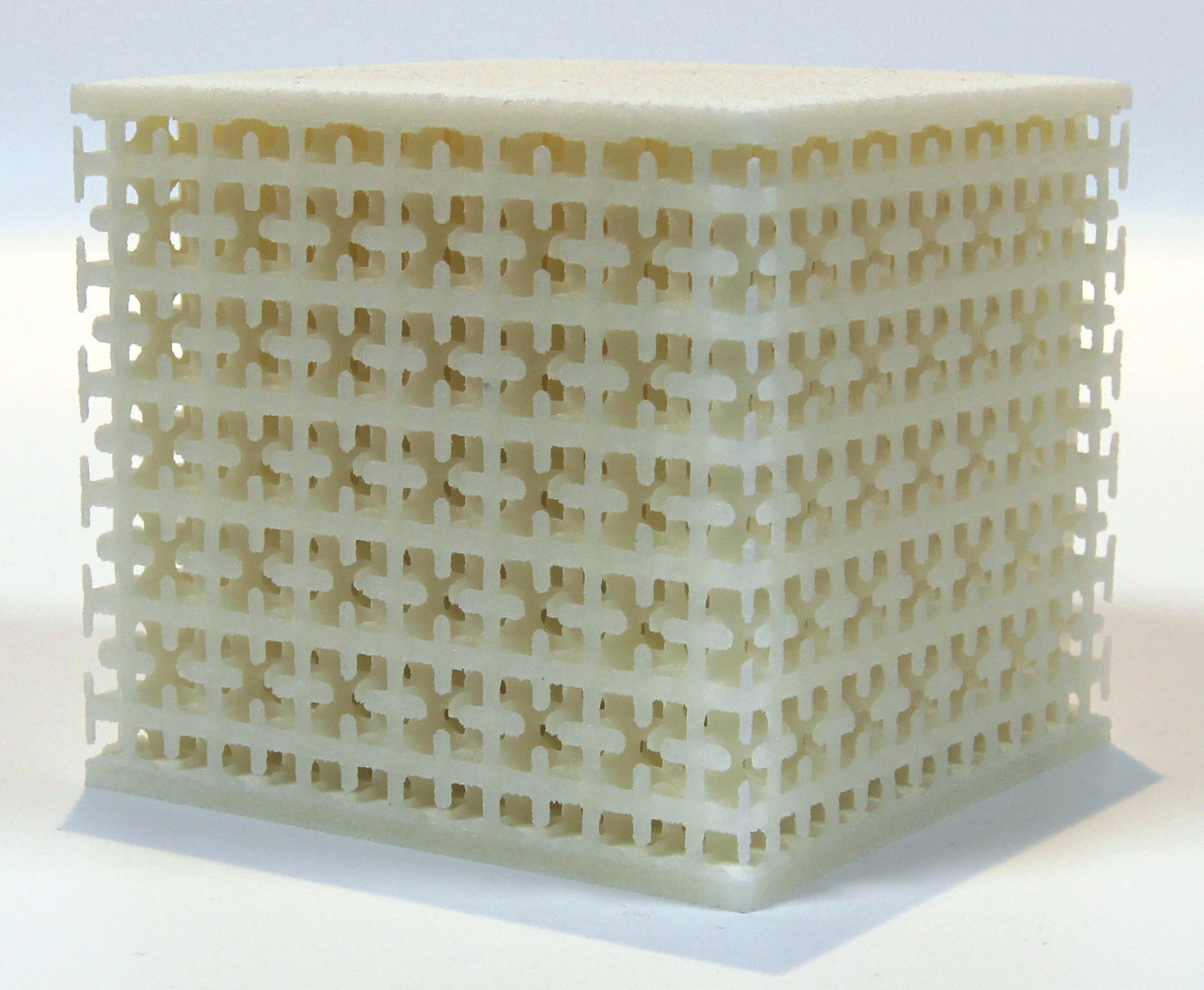}
\includegraphics[width=0.3\linewidth]{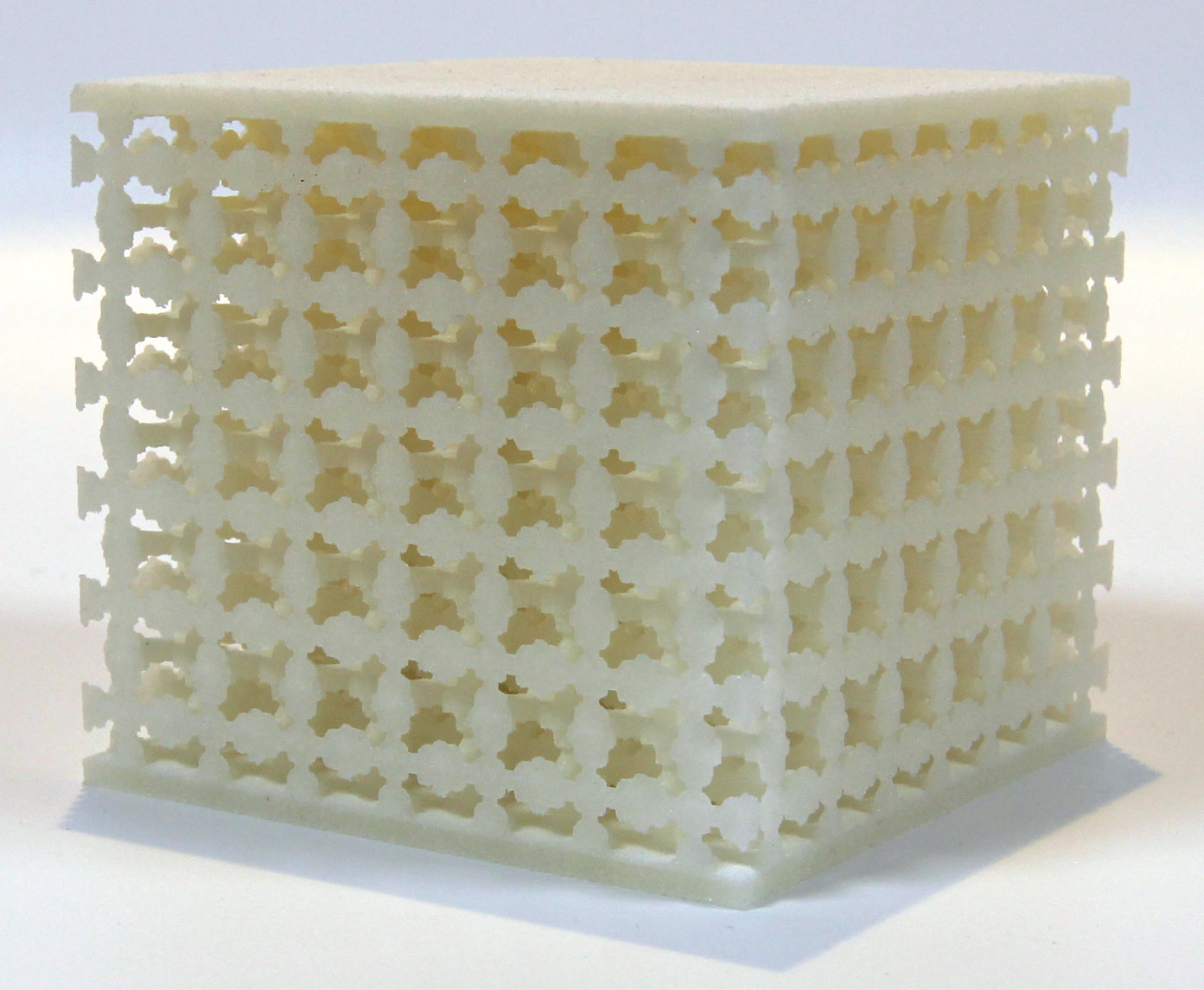}
\includegraphics[width=0.3\linewidth]{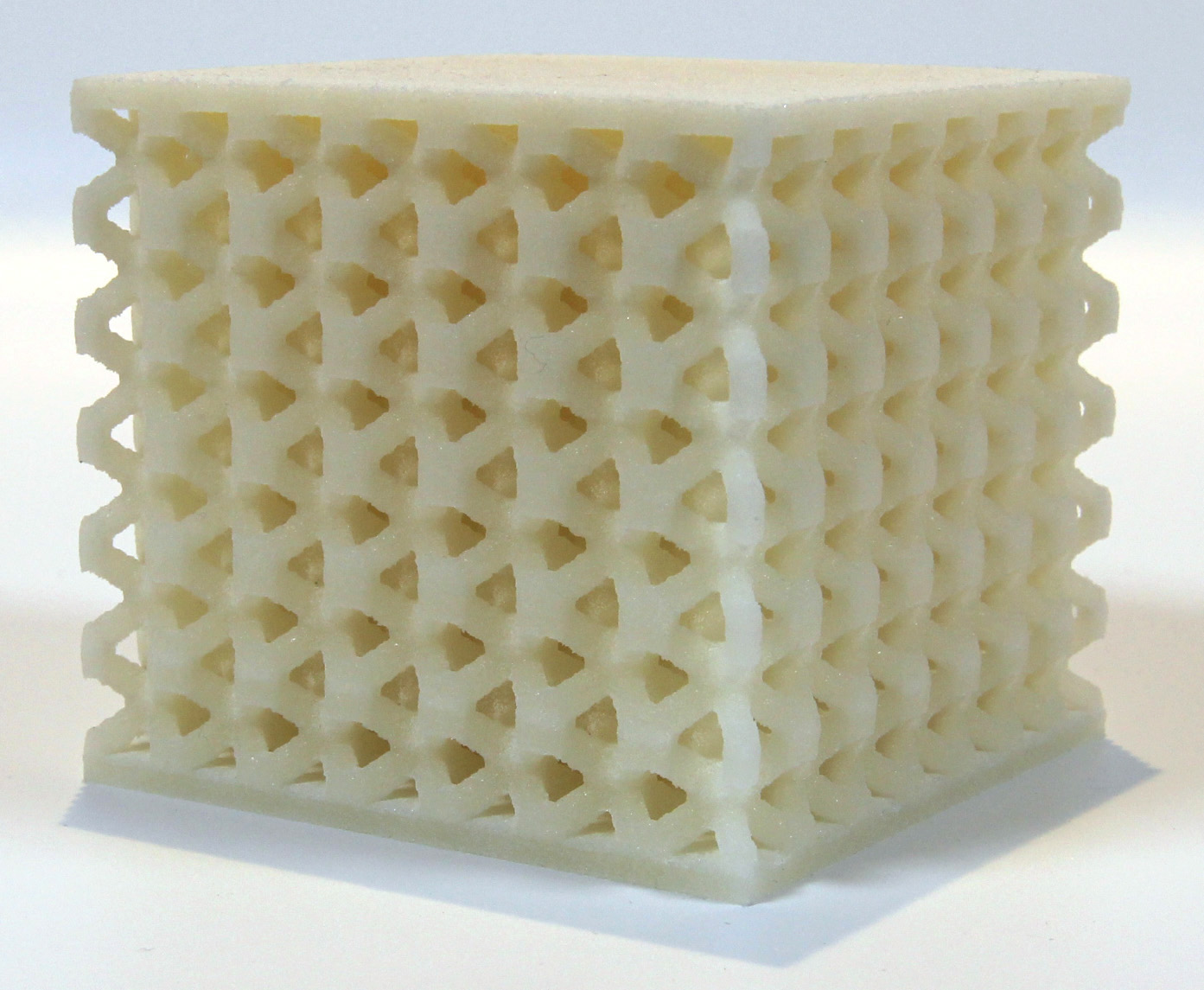}
\caption{Printed microstructures with a precomputed elastic behavior. Image courtesy of~\cite{Schumacher:2015:SBR*}}\label{fig:micro_elastic}
\end{figure}

Different works seek to produce microstructures that can be fabricated~\cite{zhou2008microstructural,Andreassen:2014:ALO} and produce a prescribed elasticity.
\cite{Schumacher:2015:SBR*} and \cite{Panetta:2015:ETA} consider periodic tilings of precomputed microstructures that cover a large spectrum of elastic behaviours (see Figure~\ref{fig:micro_elastic}).
\cite{Martinez:2016:foams} consider procedural Voronoi-based microstructures, that can be fabricated with SLA/SLS.

Micro-structures are a very promising field of research, with many potential applications~\cite{raney2015}. 
We envision that efficient techniques for designing and fabricating micro-structures will tightly integrate with process planning.


\subsection{Slicing}
\label{sec::slicing}

Slicing is central to the process planning pipeline, as it is the step where the 3D geometry is divided into a set of planar contours. These contours will be later manufactured by material deposition. In the following we assume that the build direction is along the z axis, aligned with the {height} of the object. Each slice is a plane intersecting the shape at a given height. Assuming the shape is a solid, then its intersection with a slice plane is a closed 2D contour.

There are two important questions to solve when considering slicing: how to determine the set of slices to use and their vertical position, and how to efficiently compute the contour within each slice given the input shape and the set of slices. The two following sections discuss each of these steps.

\subsubsection{Uniform and adaptive slicing}
\label{sec:slicing_unif_adapt}
\begin{figure}[tb]
	\centering
	\includegraphics[width=.95\linewidth]{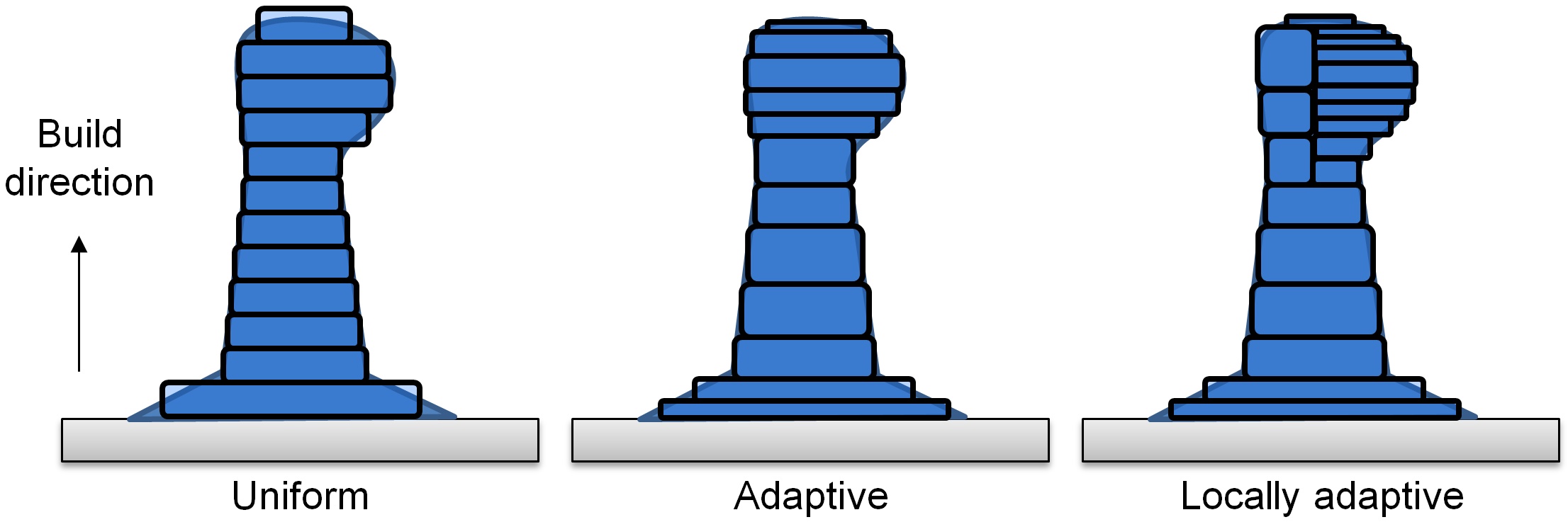}
	\caption{A same object sliced with different approaches. \textbf{Left:} uniform slicing (12 slices), \textbf{middle:} adaptive slicing (12 slices, volume error is reduced), \textbf{right:} locally adaptive slicing, where the top region is split into two sub-regions sliced independently.}
	\label{fig:slicing}
\end{figure}

The simplest approach to divide the object into slices is to subdivide it uniformly, as illustrated in Figure~\ref{fig:slicing}, left. Given a manufacturing layer thickness $\tau$ and an object height $H$, the object is divided into $N = \lceil \frac{H}{\tau} \rceil$ slices. Each slice $i$ is then located at height $z_i = \frac{i+0.5}{N}$, which is the position of the plane that will be intersected with the object.
This approach is widely adopted and most software offer it as a standard approach.

However, many objects have a shape that varies greatly along its height. Therefore, in some regions uniform slicing might use too many slices, while it does not properly capture the shape in others. In particular, surfaces that are slanted with respect to the build direction produce a staircase defect through additive manufacturing. These areas require to use very thin slices (small $\tau$). Uniform slicing forces the same small value of $\tau$ to be used throughout the entire part, which produce a large number of slices and increases manufacturing time. 

Most technologies are able to change the layer height during manufacturing. \textit{Adaptive slicing} approaches exploit this property, by adapting the thickness of each slice to the shape geometry, as illustrated in Figure~\ref{fig:slicing}, middle. Given a model for the geometric error (see Section~\ref{sec:desiderata:fidelity}), these approaches refine or coarsen the slices to meet a quality constraint while reducing print time. This can be achieved by locally determining the slice thickness from the error \cite{dolenc1994slicing,Suh:1994:ASO}, by subdividing the slices from the coarsest uniform set of slices \cite{Sabourin:1996:ASU,Kulkarni:1996:AAS,Hope:1997:ASW}, by merging slices starting from the thinnest uniform slices \cite{Hayasi:2013:ANA}, or by formulating a global optimization problem \cite{Wang:2015:SPS,Sikder:2015:GAS}. 

While adaptive slicing is able to adapt to shape changes along the build direction, it still cannot adapt to a change in part complexity \textit{within} the layer. Consider an object with a vertical wall on the left, and a slanted surface on the right. The vertical wall could be printed with thick slices, however the slope on the right imposes the use of thin slices to limit the staircase effect. To reduce this issue, \textit{locally adaptive slicing} has been proposed \cite{Tyberg:1998:LAS}. The key idea is to first subdivide the object into different regions, each region being sliced independently, as is illustrated in Figure~\ref{fig:slicing}, right. Depending on the target technology, the different regions can be built together by locally changing the layer height \cite{Tyberg:1998:LAS,Wang:2015:SPS}, or they can be printed independently and later assembled \cite{hildebrand2013orthogonal,Wang:cgf:improved}.
The main issue with this approach is that seams appear along the surface where different layer thicknesses meet. 
For techniques where parts are printed separately a manual assembly step is required.

Interestingly a similar approach was used on the object interior by \cite{Sabourin:1997:AEF}: Since the inside is never visible, it can be sliced using a larger thickness than the exterior.
This idea can be combined with the aforementioned techniques, for instance performing local adaptive slicing only on the exterior shell while the interior uses the maximal thickness~\cite{Mani:1999:RBA}.

\subsubsection{Slice contouring}

Once the set of slices determined, each slice plane has to be intersected with the input geometry. This operation strongly depends on the representation of the input. 
We first discuss slicing of triangle meshes and ray-representations (ray-reps). Both are well studied and successful approaches, which are often described as \textit{indirect} approaches : the input CAD model has to be converted into a triangle mesh (tessellated) or a ray-rep (ray-tracing or rasterization). Both conversions require the user to set a precision parameter and may loose information. We discuss these issues in more details below.
Therefore, a number of \textit{direct} slicing techniques have been proposed, that avoid any re-sampling of the initial CAD model. We discuss these contouring techniques last.

\paragraph{Contouring triangle meshes} 

A general scheme for contouring triangle meshes consists in first extracting all intersection segments between the slice plane and the triangles, and then forming loops \cite{Kirschman:1992:APS}. 
If the input correctly defines a non self-intersecting solid, the loops will be closed and non-intersecting. Otherwise, a mesh repair step is required (see
Section~\ref{sec::repairing}). Alternatively the slicer may attempt to close holes between nearby segments and resolve intersections.

Implementations mainly differ by how segments and loops are formed. Kirschman and Jara-Almonte \shortcite{Kirschman:1992:APS} propose a parallel implementation that intersects each slice plane with all triangles. McMains and S\'equin \shortcite{McMains:1999:ACS} propose an efficient algorithm that exploits the mesh connectivity to sweep a slicing plane through the triangles. The observation is that the topology of the 2D contours remains the same between vertices of the input mesh. Therefore, the contours can be very efficiently produced for all slicing planes in between two vertices. The update to be performed at each mesh vertex is often limited and fast.
The open source software \textit{CuraEngine} traverses triangles first, and each is intersected by the slicing planes it covers (see \texttt{Slicer::Slicer} and \texttt{project2D} in \texttt{slicer.cpp/.h}). The segments are identified by the faces to which they belong, and contours are formed by looping over segments following mesh connectivity (see \texttt{makeBasicPolygonLoop} in \texttt{slicer.cpp}).
Zhang and Joshi \shortcite{Zhang:2015:AIS} argue that mesh connectivity might be expensive to obtain and propose to incrementally construct the contours while triangles are traversed. Linked lists of segments are augmented by adding the new segment to the head or tail (or starting a new list). The segments are not identified directly by intersection points, but instead by the triangle edge to which they belong -- making the approach robust to numerical errors in intersection computations. 

\paragraph{Contouring~of~ray-representations}
\label{sec::slicing:rayrep}
\noindent Ray-reps are techniques were the geometry is captured by solid/empty intervals along a set of rays. 
This technique was pioneered by Hook~\shortcite{Hook:1986:RTS} who proposed the \textit{dexel-buffer}. This data-structure is built by intersecting axis-aligned rays along one direction with the geometry. Given a closed geometry, the number of intersections is even and each interval can be classified as inside (solid) or outside (empty).
The dexel-buffer is closely related to the A-buffer~\cite{Carpenter:1984:BAH}. There are therefore interesting similarities between ray-reps and A-buffer techniques for order independent transparency~\cite{maule.11.cag}. For instance, dexel-buffers can be constructed by rasterization, recording all fragments drawn in every pixel~\cite{Lefebvre:2014:PPL}. Other efficient approaches are based on Layered depth images (LDI)~\cite{Shade:1998:LDI}, which capture the geometry as a set of depth images each storing a surface sheet in space (a \textit{depth peel}). This data structure has been extended to solid modelling for additive manufacturing by Chen et al.~\shortcite{Chen:2008:LDN,Chen:2008:LDN}, adding normal information along with depth information (layer depth \textit{normal} images, or LDNI). Fast construction methods have been proposed, e.g. using single pass voxelization~\cite{Leung:2013:CSO} and compaction for increase memory efficiency~\cite{Zhao:2011:PAE}. Most ray-reps used multiple directions to better reproduce geometries~\cite{Benouamer:1997:BTG}.

\begin{figure}[tb]
\centering
\includegraphics[width=\linewidth]{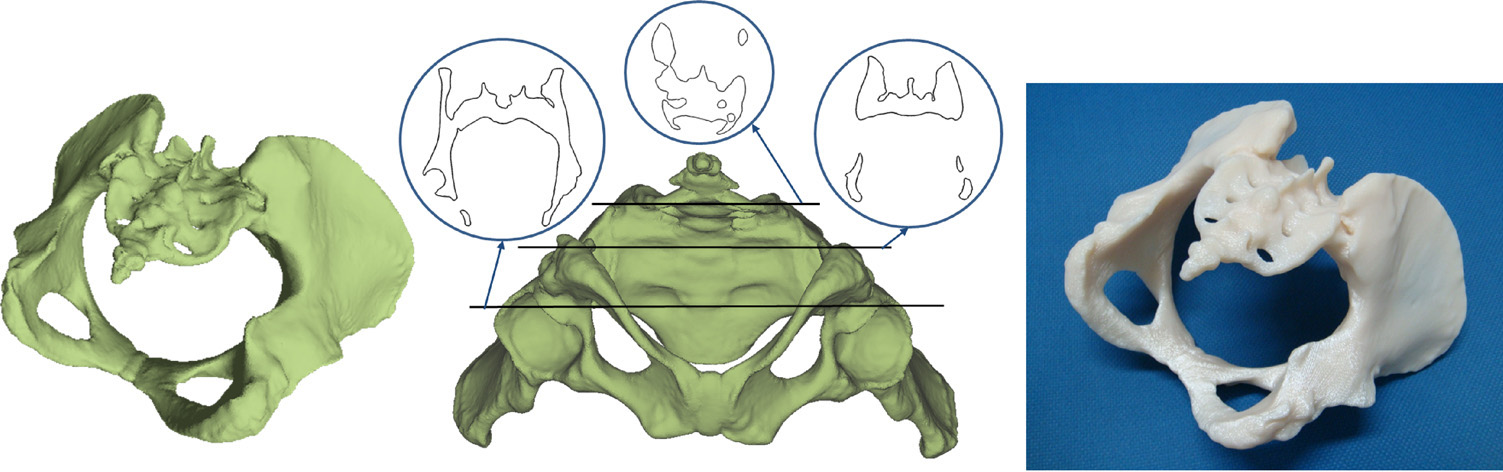}{}
\caption{Slicing and contouring of a ray representation LDNI while preserving its topology. Image courtesy of~\cite{Huang:2013:IFA}.} \label{fig:faithful_slicing}
\end{figure}

Once obtained, ray-reps may be directly rendered~\cite{Hook:1986:RTS,Wang:2010:SMO,lefebvre.13.aefa} or converted into meshes through efficient procedures~\cite{Zhang:2009:SRU,Wang:2010:SMO,Wang:2011:ABO}. However, they may also be directly contoured to extract slices. Different strategies have been used: single set of rays from the object side~\cite{Zhu:2001:DBD,Zeng:2011:ESP}, rays from two sides~\cite{Qi:2013:RSP}, and rays from the object bottom~\cite{lefebvre.13.aefa}. Contouring is then performed by marching along the rays, forming polygonal loops~\cite{Zhang:2007:ANC,Yuksek:2008:CAN}. A key issue when using ray-reps for contouring is to decide upon the resolution required to properly capture the geometry and its topology~\cite{Huang:2013:IFA} (see Figure~\ref{fig:faithful_slicing}).
The contours extracted from ray-reps have typically many small segments -- each slice is an image and contours are extracted as the outline of the solid pixels. Huang et al.~\shortcite{Huang:2013:IFA} describe a topology preserving contour simplification, and a full image based pipeline for additive manufacturing~\cite{huang.chapter.14}.
Another difficulty is the large memory requirements, which is roughly proportional to the surface area. To avoid saturating memory, tiling schemes may be used~\cite{Chen:2013:RCG}. 

Ray-reps have other significant advantages for additive manufacturing, for instance to perform CSG between complex geometries~\cite{Benouamer:1997:BTG,Wang:2010:SMO,lefebvre.13.aefa}, for regulating solids having self-intersections~\cite{Chen:2013:RCG} or for computing offset surface for hollowing parts (detailed in Section~\ref{sec:sup:internal:hollow}). 

\paragraph{Direct slicing}
\label{sec:slicing:direct}
To avoid having to re-sample the CAD model into a triangle mesh or ray-rep, several techniques have been proposed to extract contours directly from the initial geometry. These approaches directly output slice data to the printers. The slice file format is often vendor-dependent, but some independent formats such as CLI (Common Layer Interface) can be used. Open source printers often accept slice data (e.g. G-Code for filament printers or images for DLP printers such as the B9Creator or Autodesk Ember). 
Jamieson and Hacker~\shortcite{Jamieson:1995:DSO} provide an in-depth discussion of the pros and cons of direct slicing of CAD models for different input types. 

Non-uniform B-spline surface (NURBS) are a common surface representation in CAD software. Therefore, approaches have been proposed for the direct slicing of NURBS models in order to avoid a global tessellation of the geometry~\cite{Vuyyuru:1992:RP}. Later approaches consider specialized adaptive slicing~\cite{Ma:2004:NBA} and orientation procedures~\cite{Starly:2005:DSO}.

\begin{figure}[tb]
\centering
\includegraphics[width=\linewidth]{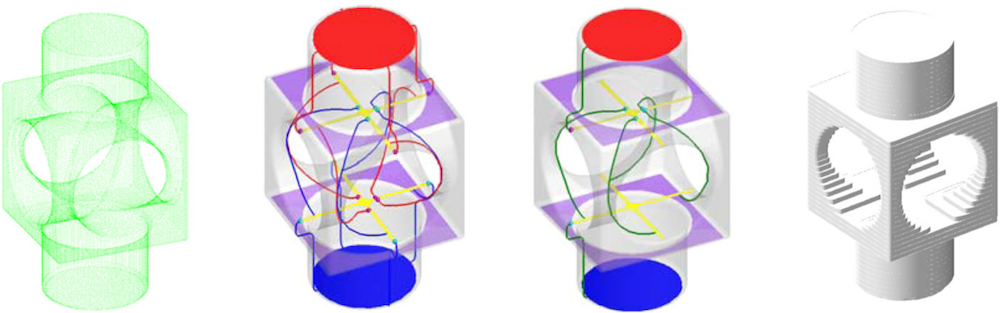}{}
\caption{Direct slicing (right) of a point cloud (left) via topological analysis (middle). Image courtesy of~\cite{Yang:2010:TES}.} \label{fig:topo_slicing}
\end{figure}

Techniques have also been proposed for point clouds, which are often obtained from 3D scanners or vision algorithm. They are challenging to print since the connectivity and topology of the surface is unknown.
Early methods project points around the 2D slicing plane and reconstruct a contour~\cite{Liu:2003:EBS,Wu:2004:MCD,Shin:2004:DSO}. Yang and Qiang~\cite{Yang:2008:ASO} propose to rely on the moving least square method to implicitly define the surface from the point cloud. Qiu et al.~\shortcite{qiu.11.ijamt} and Yang et al.~\cite{Yang:2010:TES} refine this approach by considering the global topology of the shape to detect and capture extremal points in an adaptive slicing strategy (see Figure~\ref{fig:topo_slicing}). Chen et al.~\cite{Chen:2013:DGP} propose a complete system for scanning an object and fabricating it in a different location, based on point clouds. This includes a novel support generation algorithm. 

\paragraph{Non-planar approaches}
\label{sec:slicing:alternatives}

There has been a number of interesting attempts at moving beyond standard layers for additive fabrication. These attempts usually
focus on a specific technology since they exploit properties such as the ability to perform Z-motion during deposition (fused
filament fabrication) or the ability to partially cure material (SLA/DLP). 

In the context of stereo-lithography (SLA) Pan et al.~\cite{Pan:2012:SSF,Pan:2015:SSF} exploit the formation of a meniscus when an object moves out of the resin tank. The meniscus is cured to fill the creases between two layers. Repeating this process produces smooth, accurate surfaces. Park et al.~\cite{Park:2011:DMF} show how dithering can cure resin partially and produce slanted surfaces along a layer.

In the context of filament deposition, Chakraborty et al.~\shortcite{Chakraborty:2008:EPG} proposed study curved layer deposition
with the objective of strengthening shell-like parts by aligning the toolpaths with the surface. 
The mechanical properties of the parts are discussed in~\cite{Singamneni:2012:MAE}.
An interesting question is then to combine flat and curved layers, as  discussed in~\cite{Huang:2012:ASA,Allen:2015:AED}. 
A key challenge of curved layers deposition is that the curvature of the paths is large as they flow along the surface several millimetres up and down. This has been demonstrated for specific parts, however in a general setting this degree of freedom is challenging to exploit. First because
a novel type of slicers have to be developed, and second because the current design of deposition nozzle complicates the task: collisions between already printed paths and currently deposited paths become possible~\cite{Chakraborty:2008:EPG,Allen:2015:AED}.

\begin{figure}[tb]
\centering
\includegraphics[width=\linewidth]{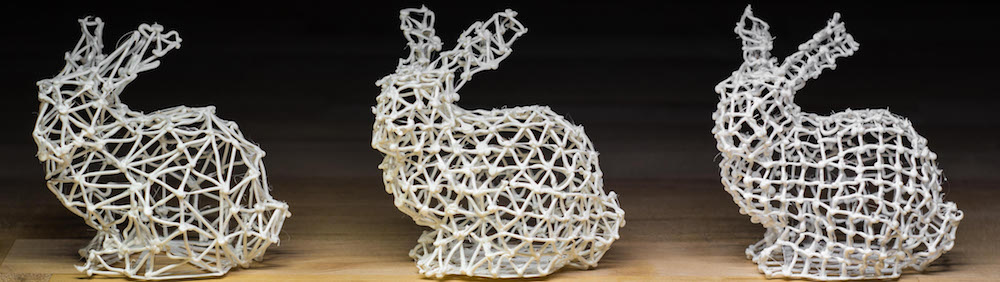}
\caption{Printed 3D wireframes, using a 5DOF printer. Image courtesy of~\cite{Wu:2016:PAM}.} \label{fig:wire_printing}
\end{figure}

Beam and truss structures are not very well suited for layer by layer fabrication, as the beams are sliced into many small cross-sections. Mueller et al.~\shortcite{Mueller:2014:WDP} propose a dedicated approach to print wire-mesh structures, exploiting the fact that extruded filament hardens quickly to print truss-like structures in mid-air. This raises many challenges regarding path planning of the extrusion device. Wu et al.~\shortcite{Wu:2016:PAM} propose a more general approach using a 5DOF printer (see Figure~\ref{fig:wire_printing}) while Huang et al.~\shortcite{Huang.16.tog} rely on a robotic arm. Both propose a path planning algorithm avoiding collisions.

\paragraph{Multiple materials.}
\label{sec:multimat}
Some technologies afford for multi-material fabrication. 
Weiss et al.~\shortcite{Weiss:1997:SDM} describe a process for fabricating multi-material objects.
Kumar et al. \cite{Kumar:1998:AAT,Kumar:1998:slicing} describe a modeling representation and adaptive slicing algorithm for multi-material objects. 
Zhu and Yu~\shortcite{Zhu:2001:DBD} propose a dexel slicer for multi-material objects. The solid ray intervals are used as solidification paths. 
Park et al.~\cite{Park:2001:VMT} describe a system for modeling and fabrication of multi-material objects inspired by procedural volume texturing approaches in computer graphics, noting the advantages for the process planning in terms of memory compactness and resolution independence. 
Shin et al.~\cite{Shin:2003:MDF} focuses on process planning for fabricating objects with continuously varying material properties on a direct metal deposition system.

\begin{figure}[b]
\centering
\includegraphics[height=3.1cm]{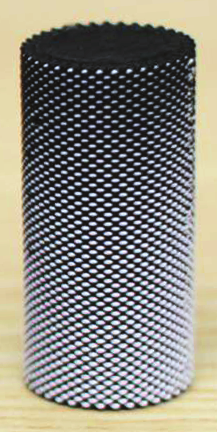}
\includegraphics[height=3.1cm]{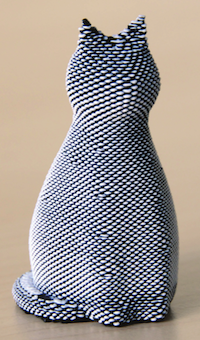}
\includegraphics[height=3.1cm]{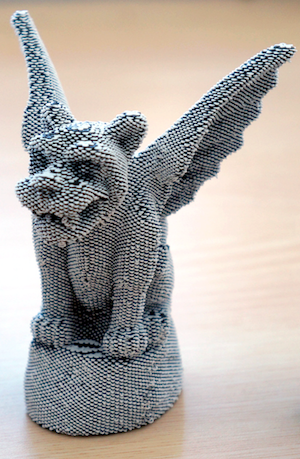}
\includegraphics[height=3.1cm]{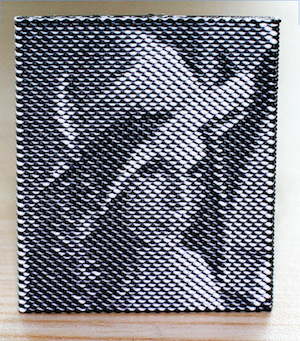}
\caption{Producing continuous tone imagery by color mixing on multi-filament printers. Image courtesy of~\cite{Reiner:2014:DCM}.}\label{fig:color_mixing}
\end{figure}

Machines based on filament extrusion can mount multiple extrusion heads, in which case motion planning during deposition becomes more complex. Choi and Cheung~\shortcite{Choi:2005:AMM} input a triangle mesh per material (packed in a single colored STL file) that are sliced independently. The contours in each layer are grouped by inclusion order, so as to reduce redundant motions during deposition. This is later extended to motion planning for multiple, independent deposition devices. The challenge is to coordinate the movements of the nozzle depositing different materials, so as to avoid collisions or even deadlocks \cite{choi2006topological,choi2010dynamic}.
In the context of computer graphics, Hergel and Lefebvre~\shortcite{Hergel:2014:CCI} also consider the case of slicing and toolpath planning for multiple materials. Instead of inputing multiple meshes, materials are selected by a \textit{slice shader} which is executed on every point of each layer. Contours are extracted using an image space approach (see Section~\ref{sec::slicing:rayrep}). The motion planner is optimized for visual quality, hiding potential defects in less visible regions of the part.
Reiner et al.~\cite{Reiner:2014:DCM} achieve a visual grading of colors on multi-filament printers, interleaving the deposited filaments in sine wave patterns along the surface (see Figure~\ref{fig:color_mixing}).

Some technologies apply colors on entire layers (inkjet on powder~\cite{zcorp} and inkjet on laminated paper~\cite{mcor}). At slicing time, a raster RGB color image is applied to the layer contour. Since the colors are often specified from the surface (e.g. vertex colors or 2D texture map), they are propagated inside from the contour within a thin shell. However, the technology would be able to color the entire volume, even though opacity of the material limits the potential.

Multi-jet technologies, the deposit droplets of different resins -- have a wider variety of materials and colors.  
OpenFAB~\cite{Vidimce:2013:OPP} allows to models complex multi-material geometries with voxels, and streams slices to a high-resolution multi-jet printer. 
This forms the basis of a complete modeling system for multi-material modeling~\cite{Vidimce:2016:uist}. 
Wu and colleagues~\cite{Wu:2000:MFD,Cho:2001:MFD} discuss process planing to convert continuously varying material information into a limited set of base materials by a half-toning technique. 
Brunton et al.~\cite{Brunton:2015:PTL} also rely on half-toning in the context of color reproduction -- to the point that print-out of scanned objects can be confused with their model. 
Earlier work also studied how to control optical properties of the final print by combining multiple base materials~\cite{Hasan:2010:PRO,Dong:2010:FSV} -- even though these approaches are not integrated within the process planning pipeline.







\subsection{Machine instructions}
\label{sec:mach_instr}


When talking about machine instructions it is necessary to distinguish between machines that operate on each slice like a
plotter, that is, connecting pairs of points with straight lines, and machines that operate on each slice like an inkjet printer, that is, interpreting the whole slice as a discrete 2D image. The former require the slices to be defined in vector format, whereas the latter require the slices to be defined in raster format (see also Section~\ref{sec:technologies}).

%

%
%
%

\subsubsection{Vector case}
\label{sec:mach_instr_vec}
\begin{figure}[h]
\centering
 \includegraphics[width=\linewidth]{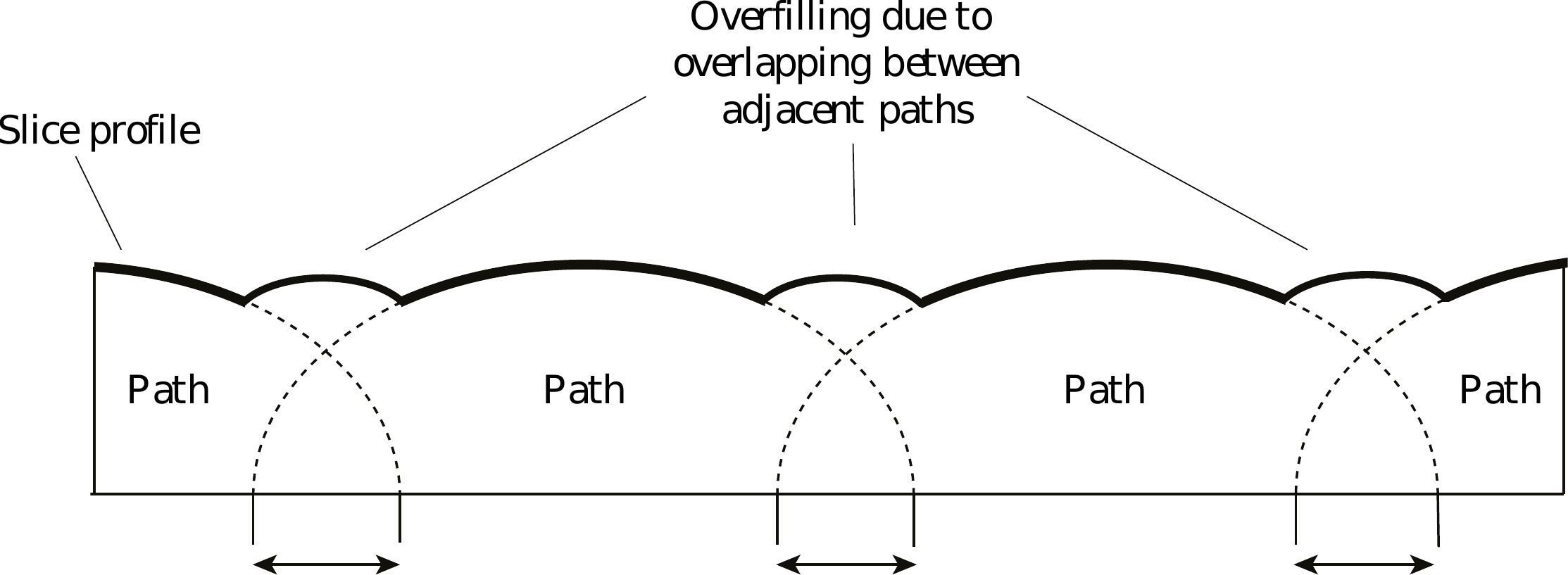} 
\caption{RP forming principle.}
\label{fig:forming_principle}
\end{figure}

In the vector case the machine instructions amount to a piecewise linear toolpath (in some cases arcs) along which the printer must deposit, melt or sinter the printing material. The machine toolpath must be prepared for each slice, both for its outer contour and the interior. If the part comes in the form of a boundary representation (e.g. a STL file), particular attention must be paid to distinguish between the interior of and the exterior of the shape \cite{volpato2013identifying}. The generation of a machine toolpath for AM has a clear analogy with CNC pocket milling, where the material inside an arbitrarily closed boundary on a flat surface is removed to a fixed path \cite{makhanov2007advanced,held1991geometry}. In the general case a machine toolpath intended for milling is, however, not suitable for additive manufacturing. In FDM the deposition of material along the path poses some additional challenges. The path has to be designed in such a way that the deposition of material is as regular as possible, thus avoiding under and over deposition. For the same reason the overlap between adjacent paths needs a much finer control, in order to cover the slice area with a uniform layer of deposited material \cite{han2003process} (Figure~\ref{fig:forming_principle}). In powder bed technologies such as laser sintering and melting the heat control is fundamental to guarantee quality results. Many printers generate multiple melting pools at a time, in order to better distribute heat and avoid huge thermal gradients, that would generate rough surfaces and warping \cite{ding2014tool}. We discuss here the requirements for a quality machine toolpath and the major differences between available algorithms and technologies.

\paragraph{Continuity} In FDM and other material deposition processes it is important to keep the amount of material deposited along the path constant. To this end, many authors observed that controlling the amount of material being deposited when a path begins or ends is very difficult. Recent works aim to reduce as much as possible the number of disconnected paths which are necessary to cover the slice area, so as to minimize discontinuities in the deposition process \cite{Zhao:2016:CFS}. 

\paragraph{Geometry} Not only the endpoints of a machine path but also the geometry of the path itself may affect the quantity of material locally deposited in FDM. Long and low curvature paths are to be preferred to short paths with sharp turns. In the latter case the speed of the nozzle would decrease, thus increasing the time necessary to complete the deposition and possibly triggering under or overfilling of the filament \cite{jin2014optimization}.

\paragraph{Patterns} 
Internal volumes of parts are a major factor in time and material consumption, in particular for FDM where the motion of the print head -- a relatively heavy mechanical device -- is slow. To save time and material, several infill patterns have been proposed. These patterns are usually specific to FDM due to the aforementioned continuity and geometry requirement, but also due to the overhang constraints. 
See Section~\ref{sec:sup:internal:infills:sparse} for an in-depth discussion of sparse infill patterns.

\paragraph{Domain split} In order to be able to process arbitrarily complex slices, most of the approaches proposed in the literature use a divide-and-conquer strategy, partitioning the slice into a set of pockets to operate on, rather than trying to cover the whole area with a single connected curve. To this end, many different strategies have been proposed in the literature. In \cite{dwivedi2004automated} the slice is decomposed into a set of monotone polygons; \cite{ding2014tool} uses convex decomposition; \cite{Zhao:2016:CFS} uses the iso-contours of an inward distance field from the border; \cite{held1994pocket} uses a Voronoi-based approach and \cite{kao1998optimal} uses a medial axis based approach.

\paragraph{Performances} Besides the method used to decompose the domain and the particular curves used to fill each pocket, another important factor in the definition of an efficient toolpath is the reduction of the so called machine \emph{airtime}, that is, the time necessary to move the nozzle (or the laser) from the end of a curve to the beginning of the subsequent one. Given a toolpath composed of a set of disconnected curves, the machine airtime can be minimized by acting on to three separate variables: the order in which the curves are processed, the path that takes the nozzle from the end of a curve to the beginning of the next one, and the orientation of each curve (from begin to end, or from end to begin). For the latter, notice that for closed loops, the machine may chose any point within the loop as a starting point. The path planning optimization problem has been shown to be related to the Travelling Salesman Problem (TSP), which is NP-complete. Many methods have been proposed for its approximate solution, using genetic algorithms \cite{weidong2009optimal,wah2002tool}, the Christofides algorithm \cite{fok20163d,ganganath2016trajectory} or other heuristics \cite{castelino2003toolpath}. 


\subsubsection{Raster case} 
Recent printing technologies such as Z-Corp. and the DLP/SLA printers (i.e. stereolithography printers that exploit the Digital Light Processing technology) treat each slice as an array of pixels. For the Z-Corp. case the machine toolpath consists in a trivial visit of a regular 2D grid. From a machine usage point of view an important parameter is the binder saturation, which affects both the strength and the accuracy of a printed part \cite{vaezi2011effects}. Typically, a binder is first applied with a higher saturation to the edges of the part, creating a strong \emph{shell} for the exterior. Next, an infrastructure is created for the part walls, which are also built with a higher saturation. The remaining interior areas are printed with a lower binder saturation, which gives the part its stability \cite{z450manual}. For the DLP-based printers there is no notion of machine toolpath, because the whole image/slice is projected onto the first layer of the resin tank. In this case it is important to project each slice for the proper amount of time (exposure time), in order to balance surface accuracy with part strength. Exposure time depends on material, layer thickness and slice resolution. Notice that commercial slicers allow to export vector slices (e.g. in SVG format) even for raster printers \cite{slicer}. This is because slice rescaling is usually performed internally prior to projection, and scaling a raster image may introduce unnecessary artifacts such as image blurring.


\begin{table*}[t]
	\centering
	\begin{small}
		\begin{tabular}{| l | c | c | c | c |}
			\hline
			&\textbf{Orientation }
			&\textbf{Supports }
			&\textbf{Slicing } 
			&\textbf{Toolpath } \\
			\hline
			\textbf{Cost} &&&&\\
			$\qquad$ Pre-build				& - & - & - & Sec.~\ref{sec:mach_instr_vec} \\
			$\qquad$ Build					& Sec.~\ref{sec:orient_cost} & Sec.~\ref{sec:sup:internal}  & Sec.~\ref{sec:slicing_unif_adapt} & Sec.~\ref{sec:mach_instr_vec} \\
			$\qquad$ Post-processing	& Sec.~\ref{sec:orient:fidelity} & Sec.~\ref{sec:sup:external} & - & - \\
			\hline
			\textbf{Fidelity} &&&&\\
			$\qquad$ Form 						& Sec.~\ref{sec:orient:fidelity} & - & Sec.~\ref{sec:slicing_unif_adapt} & - \\
			$\qquad$ Texture 					& Sec.~\ref{sec:orient:fidelity} & Sec.~\ref{sec:sup:external} & - & - \\
			$\qquad$ Design compliancy	& - & -	& - 	& \\
			\hline
			\textbf{Functionality} &&&&\\
			$\qquad$ Robustness							& Sec.~\ref{sec:orient:functionality} & - & - 	&-  \\
			$\qquad$ Mass distribution					& - & Sec.~\ref{sec:sup:internal}	& 	- & - \\
			$\qquad$ Thermal/Mechanical prop.	& - & Sec.~\ref{sec:sup:internal}	& - 	& - \\
			\hline
		\end{tabular}
	\end{small}
	\caption{Relations between process planning steps and quality metrics.}\label{tab:shortcomings}
\end{table*}

\subsection{Multiple components / Batch printing}
\label{sec::batch}
In principle, a 3D model made of multiple parts can be printed all at once, that is, pre-assembled. However, with current technologies this approach might still have several drawbacks:
\begin{itemize}
\item The pre-assembled model can be too large to fit the printing chamber, whereas each single constituting part would fit;
\item The model is small enough, but it contains tangency parts which would be fused together during the printing process (e.g. this might happen for bearing spheres);
\item The model has none of the aforementioned issues, but printing the assembly would require the insertion of support structures that could not be removed;
\item The model has none of the aforementioned issues, but printing the assembly would require the insertion of support structures that, after removal, would produce rough surfaces (above the tolerances set in the design).
\end{itemize}

For all these reasons, the current practice for the manufacturing of components made of multiple parts consists in building each part separately, and thus requires a reassembly of the physical objects to form the ultimate component. Some elements of the design (e.g. screws, bolts, bearings, ...) are bought from external producers that offer a catalogue of standardized components, whereas all the remaining parts are produced by additive manufacturing. Typically, each part undergoes most of the process planning steps independently. Each so-prepared part (that is, checked, oriented, and supported) is sent to the software that drives the printer and, through this software, an operator places the part in a free portion of the building plate. When the plate is full, or when there are no more parts to print, the operator runs the actual printing process. In this case, the only process planning steps which are common to all the parts are the slicing and the creation of the toolpath. It is also possible to slice one part at a time while using the same layer thickness: this is done, for example, on EOS machines which use the PSW software to place pre-sliced parts on the building platform. Algorithmic approaches exist that try to optimize for both quality and packing efficiency by properly orienting and placing the various parts on the platform \cite{canellidis2006pre,zhang2015optimization}.

\subsubsection{Printing big objects}
\begin{figure}[tb]
\centering
\includegraphics[width=\linewidth]{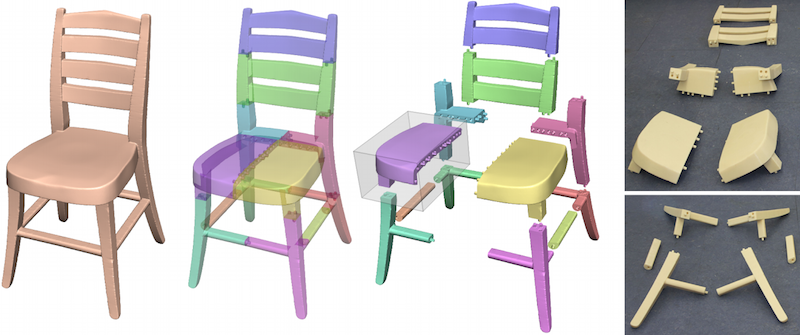}
\caption{A 3D object is partitioned into reassemblabe parts that can be printed. Image courtesy of~\cite{luo2012chopper}.} \label{fig:chopper}
\end{figure}

When even a single object is larger than the printing chamber, solutions exist to split it into parts to be printed independently and reassembled afterwards. A noticeable example is given in \cite{luo2012chopper}, where the algorithm decomposes a 3D object into reassemblable parts, each contained within a given printing volume. Besides reassemblability, this algorithm strives to avoid the production of too small parts, and considers the structural soundness by avoiding to put seams in areas of high mechanical stress (see Figure~\ref{fig:chopper}).
In a similar work \cite{hao2011curvature}, the seams are placed along lines of high curvature so as to minimize their impact on the aesthetics. In both \cite{luo2012chopper} and \cite{hao2011curvature}, connectors are created to ease the actual reassembly.
Conversely, instead of using connectors, glue, or screws, in \cite{song2015interlocking} the subdivision strives to create a part configuration that allows self-interlocking, such that the assembled object can be not only repeatedly disassembled and reassembled, but also strongly connected by the parts' own geometry.

\subsubsection{Print, pack and ship paradigm}
\label{sec:multi:ship}
Delivering a printed object (e.g. to a customer) has a cost that grows as the size of the \emph{pack} grows. Thus, it is important to investigate how a 3D model can be split into easily printable parts that can eventually be tightly packed in a box and reassembled at the destination. In \cite{attene2015shapes}, an algorithm that performs this kind of split is proposed based on a user-controllable tradeoff between the packing efficiency and the number of parts to be produced. In \cite{vanek2014packmerger}, the focus is mainly on the reduction of the material usage, for which hollowing and orientation optimization are used: however, the eventual packing efficiency is one of the parameters that drives the overall optimization process. In \cite{yao2015levelset}, the structural stresses are also considered to perform the subdivision prior to the packing. In \cite{zhou2014boxelization}, a generic shape is converted into a foldable set of nearly-cubical parts. The printer can produce the object in its folded configuration which occupies a smaller volume, and thus might fit the printing chamber. Afterwards, the printed prototype can be unfolded into the desired shape.

\subsubsection{Strive for quality prints through divide-and-conquer}
\label{sec:multi:qual}
After their removal, support structures easily leave tiny defects on the surface. To avoid this drawback, in \cite{hu_siga14} the 3D model is split into so-called \emph{approximate pyramidal shapes} that can be printed without supports. A similar approach is employed in \cite{herholz2015heightfields}, where a limited distortion is allowed to minimize the number of parts to be produced.
In \cite{hildebrand2013orthogonal} the surface quality is improved by splitting the model into few pieces so that each piece can be consistently sliced with a small geometric error along one of three orthogonal slicing directions.


\section{Relations between quality metrics and basic steps}
\label{sec::PP_analysis}

Each of the process planning stages discussed so far has an impact on some of the quality metrics. Since the relative importance of these metrics depends on the application, herewith we provide an overview of these process-quality relations that may help in the process of tuning the whole process planning. Table \ref{tab:shortcomings} summarizes these relations and, for each of the specific PP steps, refers to the corresponding section where the causes and effects are described in detail.

Note that in many cases several quality metrics should be optimized at the same time, and that is why multiple decision systems and genetic algorithms have been proposed to support the user in the difficult task of finding a tradeoff \cite{ingrassia2017process,anitha2001critical}.

\section{Open challenges}
\label{sec::issues}

In this final section we discuss some of the open challenges in the process planning pipeline for additive manufacturing.

\subsection{Challenges in an industrial perspective}
AM has a huge potential for industry, thanks to the many advantages introduced by the technology itself and also the extreme simplification of the process planning compared to traditional approaches. Nevertheless, several challenges still have to be solved to enable a large adoption of AM by the industry.

Industry typically seeks integrated solutions for PP, i.e. algorithms for model repair, part orientation, support creation, slicing, toolpath calculation etc. shall be preferably integrated in a single, comprehensive tool.
In this regard, one of the key challenges is to keep such tools up to date and allow a suitable degree of customization. At the academic level, the development of a common software framework for AM would help towards this goal: by developing their research prototypes within a common framework, academics could reach a larger audience and ease transfer towards the industry.

An interesting trend is to attempt to automatically optimize most steps of the process (part orientation, supports, batch printing, etc.). However, it is important to give enough controls to users.
For instance some users will prioritize the performance of the part (structural resistance, surface finish, etc.), while other users would be interested in optimizing the process (production time, material waste, etc.).
To this end, the parameter(s) to be optimized by PP algorithms must be selected by the user. Unlike for non-professional users, tools where the parameters cannot be tuned would not be acceptable. 

Another area of improvement of PP is the process simulation. As we have seen in this survey, some key characteristics of the final result can already be reliably forecast and some of the features can now be subjected to an optimization loop. The future development will hopefully produce more and more accurate tools that can be integrated in the PP, allowing the users to improve desiderata and minimize the uncertainties related to the AM technology.

\subsection{Complementarity with subtractive technologies}
\begin{figure}[b]
	\centering
	\includegraphics[width=.4\linewidth]{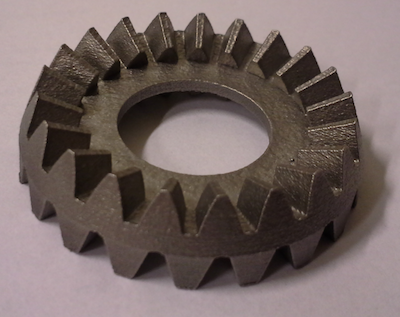} \hspace{5mm}
	\includegraphics[width=.371\linewidth]{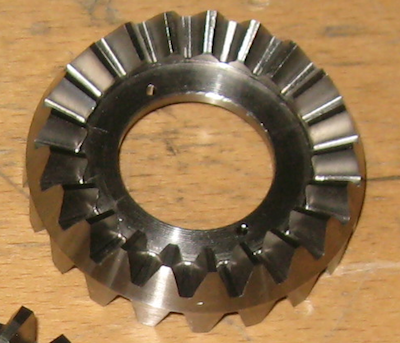} 
	\caption{Example of mechanical component (a bevel gear) manufactured with AM (left) and with subtractive technique (right). The surface finish achieved by AM is not compatible with the specifications of the gear, so a finishing operation (e.g., by milling) is required.}
	\label{fig:amvssm}
\end{figure}

In the majority of the mechanically demanding applications, AM is nowadays coupled to subtractive manufacturing. In fact, AM technologies are not capable yet of achieving the precision often required in mechanical parts, especially in terms features positioning, shape and surface finish, as illustrated in Figure~\ref{fig:amvssm}. In order to address this issue, the common practice is to modify the CAD model leaving some extra material on the features that require high precision (e.g. couplings, holes, pins, shoulders for bearings, etc.). This extra material will be then milled until the desired shape is obtained with the required surface finish. Thanks to this approach, the parts manufactured by means of AM can be used effectively and coupled to traditional parts and commercial components, with a suitable level of precision. Some recent works in process planning integrate both technologies together, e.g.~\cite{newman2015process}.

Nevertheless, the current limitation in the precision, stability and roughness of the features produced by AM is expected to be solved in the next years of technology development. Similarly to 5-axes milling, which was considerably improved with the development of CNC technologies and with the evolution of CAM tools, it is expected that soon AM will be able to produce parts perfectly meeting the stringent requirements of mechanical couplings.

This would be a double advantage, because not only it will simplify the production of parts and reduce the manufacture cost, but it would allow the users to fully exploit the freedom of AM, overcoming any constraint imposed by the finishing of features by subtractive manufacturing.


\subsection{Printing-aware modelling}
In Section~\ref{sec::pipeline} we have presented a number of techniques that, given an input shape, aim to find the best possible way to 3D print it so as to match one or a combination of the targets listed in Section~\ref{sec::desiderata}. In this section we approach the same problem from a different perspective: how can I design a shape that, if 3D printed, will satisfy the quality, cost or mechanical requirements that I would like to satisfy? The key observation is that, in many cases, tiny changes on the design would make a big difference in terms of AM. Although shape design is out of the scope of this survey, we would like to give the reader a brief overview of this topic, to emphasize that most of the problems that the process plan aims to solve, can also be tackled at design time. 
We believe that future software will couple modeling and process planning more tightly in order to guide the user towards shape the can be fabricated more reliably.

\paragraph*{Design to reduce supports.}  Support structures impact on AM in many ways (material, removal effort, surface artifacts,...). Two interesting recent works attempt to reduce supports at the design stage. 
In \cite{Hu20151} the shape is modified to make it more self-sustainable and, thus, require a minor amount of external supports. In \cite{reiner.16.egshort} the authors propose three basic 3D sculpting operations (trim, preserve and grow) to produce shapes that can be printed without supports structures. 

In general, supports are one of the most difficult and penalizing issues for AM, and more research effort is required at all stages (modeling, planning, materials and machines) to alleviate their detrimental impact.

\paragraph*{Design to ensure structural soundness.} Often, it is discovered only after fabrication that a printed shape will not be able to sustain its own weight or will be too fragile. Stava et al.~\cite{stava.12.sig} proposed not only to predict fragilities in a part, but also proposed an algorithm to reinforce the shape by means of local operations such as hollowing, thickening ans strut insertion. Although not in the context of AM, Umetani et al.~\cite{uim_guidedExploration_sigg12} proposed a design tool for furniture that suggests changes based on structural strength prediction. 
This type of feedback and automated suggestions during design could enable a widespread adoption of AM, allowing anyone to model a shape while the algorithm verifies the design and suggests fixes.

\paragraph*{Industry.} People operating in industrial design have been creating shapes to be fabricated with subtractive techniques for decades. Nowadays AM poses new challenges for designer, which have to think of new good principles that fit a different manufacturing paradigm. \emph{Printing-aware design} is often used in literature to refer to attempts to go in this direction. We point the reader to \cite{hallgren2016re,thompson2016design,salonitis2015redesign} and references therein for inspiring discussions on this topic.

\subsection{Embedding devices into prints}
A fascinating and rapidly developing topic is the ability to insert sensors or devices into a print. 
For instance, Savage et al. proposed a design tool to augment objects with specially fabricated touch sensitive areas~\shortcite{Savage:2012:MFC},
and to embed a camera into 3D printed objects to turn them into custom controllers~\shortcite{Savage:2013:SES}.
The RevoMaker~\shortcite{Gao:2015:REM} fabricates an object \textit{around} a core containing sensors and electronic devices.
The \textit{Voxel8} printer goes beyond this by proposing to integrate electronic components connected together by conductive ink deposited
within the layers.
Such conductive materials open a wide range of applications, in particular for antenna or battery designs~\cite{raney2015}.

Such applications open a whole new set of challenges overlapping between material science, geometry, 3D modeling, circuit routing, and process planning for additive manufacturing,

\subsection{Discretization-free pipeline}
As discussed earlier in this report, the usual entry point for Process Planning consists of a tessellation that approximates the smoothly curved CAD model with a collection of triangles. 
This step simplified geometric operations and exchanges between software to the point that the triangle-based STL format has become a de-facto standard for process planning. 
Nonetheless, since the tessellation step is often the cause of many issues with the geometry (see Section~\ref{sec::pipeline}) and introduces unnecessary approximations, 
many attempts have been made during the years to avoid this conversion and perform all the calculations based on the original CAD geometry (see Section~\ref{sec:slicing:direct}). 
In spite of these efforts, STL remains an undiscussed standard and CAD-compliant geometric processing for PP is still a very limited practice.

A main obstacle to a widespread adoption of such pipelines is the intrinsic hardware limitations of current printing devices. 3D printers, indeed, can just execute a finite set of commands that typically make the tool move from one point to another along a straight line segment or, in the most sophisticated models, to follow circular arcs while extruding or solidifying material. 

Nevertheless a discretization-free pipeline would allow a form of device independent process planning, where discretization would only occur at the very last moment, on the machine. Thus,
the approximations would be optimally decided depending upon machine capabilities and resolution.

\section*{Acknowledgements}
This work was partly supported by the EU project CAxMan (EU H2020 grant agreement N. 680448) and by ERC grant ShapeForge (StG-2012-307877).
Thanks are due to all the members of the Shapes and Semantics Modeling Group at IMATI-CNR and to the partners in the CAxMan project for helpful discussions.

\bibliographystyle{eg-alpha}
\bibliography{extracted}

\end{document}